\documentclass[reprint, amsmath, amssymb, aps, showkeys]{revtex4-2}

\usepackage{gensymb}
\usepackage{booktabs}
\usepackage{graphicx}
\graphicspath{{./Figures/}}
\usepackage{multirow}
\usepackage{dcolumn}
\usepackage{bm}
\usepackage{times}
\usepackage{algorithm}
\usepackage{algorithmicx}
\usepackage{algpseudocode}
\usepackage{listings}
\usepackage{natbib}
\usepackage{placeins}
\usepackage{array}
\usepackage[explicit]{titlesec}

\usepackage[colorlinks = true,
linkcolor = blue,
urlcolor  = blue,
citecolor = blue,
anchorcolor = blue]{hyperref}

\newcommand{\beginsupplement}{%
  \setcounter{table}{0}
  \renewcommand{\thetable}{S\arabic{table}}%
  \setcounter{figure}{0}
  \renewcommand{\thefigure}{S\arabic{figure}}%
  \setcounter{section}{0}
  \renewcommand{\thesection}{\arabic{section}}
  \renewcommand{\thesubsection}{\thesection.\arabic{subsection}}
  \renewcommand{\thesubsubsection}{\thesubsection.\arabic{subsubsection}}
  \setcounter{equation}{0}
  \renewcommand{\theequation}{S\arabic{equation}}%
  \setcounter{page}{1}
}

\begin{document}

\preprint{APS/123-QED}

\title{A Prompt-Engineered Large Language Model, Deep Learning Workflow \\for Materials Classification}

\author{Siyu Liu}
\author{Tongqi Wen}\email{tongqwen@hku.hk}
\author{A. S. L. Subrahmanyam Pattamatta}
\author{David J. Srolovitz}\email{srol@hku.hk}

\affiliation{Department of Mechanical Engineering, The University of Hong Kong, Hong Kong, China}

\date{\today}

\begin{abstract}
Large language models (LLMs) have demonstrated rapid progress across a wide array of domains.
Owing to the very large number of parameters and training data in LLMs, these models inherently encompass an expansive and comprehensive materials knowledge database, far exceeding the capabilities of individual researcher.
Nonetheless, devising methods to harness the knowledge embedded within LLMs for the design and discovery of novel materials remains a formidable challenge.
We introduce a general approach for addressing materials classification problems, which incorporates LLMs, prompt engineering, and deep learning.
Utilizing a dataset of metallic glasses as a case study, our methodology achieved an improvement of up to 463\% in prediction accuracy compared to conventional classification models.
These findings underscore the potential of leveraging textual knowledge generated by LLMs for materials especially in the common situation where datasets are sparse, thereby promoting innovation in materials discovery and design.

\end{abstract}

\maketitle

The concept of ``AI for Materials'' refers to the utilization of artificial intelligence (AI) techniques, such as machine learning (ML), deep learning (DL), and the increasingly prevalent large language models (LLMs), to design novel materials and to investigate composition-structure-property relationships~\cite{louie2021discovering, merchant2023scaling, li2020ai,raabe2023ncs}.
Significant strides have been achieved in fields such as biomaterials~\cite{nippa2023enabling, tropsha2023integrating} and organic materials~\cite{weiss2023guided, noe2020machine} using such approaches, which can be attributed to the relatively simple representation of the structures of organic molecules~\cite{wigh2022review} and the availability of numerous high-quality datasets~\cite{choudhary2022recent}.
However, for inorganic materials, particularly those with large compositional space (element types $>$ 4), the application of AI presents a more complex challenge.
This complexity arises due to factors such as the scarcity of experimental data~\cite{raabe2023ncs}, the diversity of material properties of interest, the complexity of crystal structures, the existence of multiple alloy phases, and the complexity/properties of defects~\cite{antoniuk2023predicting, noh2020machine} (e.g.,  vacancies, dislocations, grain boundaries).
In contrast with organic molecules that can be naturally modeled using neural networks, multi-component inorganic materials often involve the amalgamation of multiple phases and defects or even the formation of new phases, presenting an ongoing challenge for structural representation~\cite{xiao2023invertible, steinberger2019machine}.
Although large-scale databases such as the Materials Project~\cite{mp2013aplm}, ICSD~\cite{icsd2019jac}, and AFLOW~\cite{aflow2012cms} exist, data pertinent to specific tasks such as the relationships between composition and mechanical properties in alloys and structural stability of two-dimensional materials are often sparse and dispersed across different datasets, complicating the assembly of large scale training sets.
Input features often vary between datasets, necessitating manual feature construction based on domain knowledge~\cite{vu2023towards}.
Collectively, these issues have hindered AI applications in inorganic materials, subsequently impeding the discovery and design of new materials.

\begin{figure*}[t]
\centering
\includegraphics[width=0.95\textwidth]{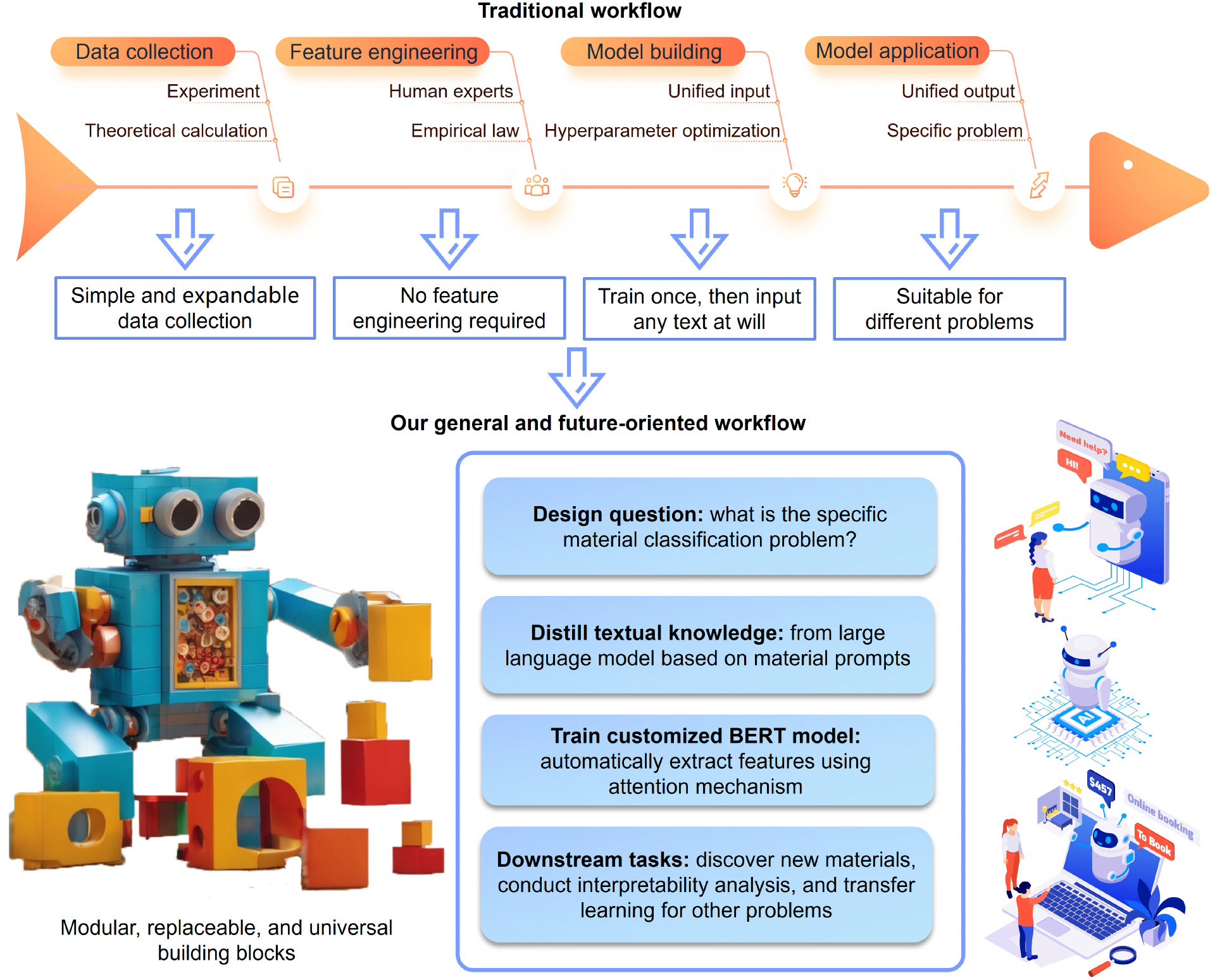}
\caption{Comparison between traditional and future-oriented general deep learning workflows. The general workflow consists of generating textual data from prompt-engineered LLMs and training deep learning models.}\label{fig1}
\end{figure*}

The advancement of LLMs, particularly \emph{ChatGPT}, sparked a surge of interest in distilling knowledge through prompt engineering~\cite{thirunavukarasu2023large, kasneci2023chatgpt, romera2023mathematical}.
Notable examples include constructing scientific question-answering knowledge bases using LLMs~\cite{pereira2023visconde}, transferring knowledge from large to small models via in-context learning~\cite{chen2024grimoire}, and training materials-specific LLMs~\cite{xie2023large, zheng2023chatgpt}.
Given the expanding capabilities of LLMs and the growing volume of training data, it is conceivable to view these models as encyclopedic resources for materials science, enabling the extraction of text-based knowledge~\cite{min2023recent}.
As shown in the upper part of Fig.~\ref{fig1}, this approach addresses some challenges associated with data collection and feature extraction in classical ML processes.
By representing content in textual format, the generated data becomes universally applicable, offering a versatile tool for materials science research.

Here, we establish a universal workflow for material text feature-label classification using textual data generated by LLMs, as illustrated in the lower part of Fig.~\ref{fig1}.
The workflow is structured as follows: (i) define the material classification problem to be addressed; (ii) design prompts (via prompt engineering) to distill knowledge from LLMs and store the information as textual data; (iii) fine-tune a bidirectional encoder representations from transformers (BERT)~\cite{devlin2018bert} model (commonly employed in natural language processing) to train on the stored textual data-label pairs; (iv) apply the model to explore new materials or study composition-structure-property relationships. 
In an example problem involving the classification of 5,577 metallic glasses (MGs) with experimental datasets labeling  MG-formability categories (bulk MG, ribbon, or non-ribbon -- i.e., glass formation under different cooling rates), a BERT model trained with optimized prompts and position embedding methods achieved up to a 48\% increase in classification accuracy compared to traditional ML models. 
The classification models obtained from the workflow presented here achieved an overall accuracy of 97.7\%.
For the classification of bulk MGs (BMGs - those formed at the slowest cooling rates) from the smallest training dataset ($\sim$11\% of the entire dataset), the accuracy improved by up to 463\% compared to traditional ML models.
These results underscore the superiority of the proposed workflow in addressing material classification problems.
Since this workflow is extendable to a wide range of material applications, this approach demonstrates the tremendous potential of ``AI for Materials'' grounded in natural language processing and prompt engineering for materials engineering.

\section*{Text-based materials classification workflow}
Our material classification workflow has four tiers depicted in the blue box in Fig.~\ref{fig1}, and is readily adaptable to various material classification problems.
Initially, researchers undertake fundamental data collection and delineate the prediction objectives and categories, such as determining whether a material exhibits stability, corrosion resistance, or favorable optoelectronic transmission properties.

The inputs to conventional ML model workflows typically comprise material properties (obtained through experiments or theoretical calculations), demanding substantial expertise and effort to obtain experimental or simulation data.
Different researchers may focus on different input features, leading to multiple datasets for the same problem~\cite{morgan2020opportunities}.
Training numerical-based ML models on a combination of different datasets often fails due to inconsistencies in input feature vectors.
The size of datasets for specific niche tasks in materials science research tends to be relatively small ($<$ several hundred entries); this is negligible compared to the millions of samples in classical computer science fields like the ImageNet dataset~\cite{deng2009imagenet}.
For example, in the field of alloys, a typical set of data for the fracture and impact toughness of high-entropy alloys contains only 154 entries~\cite{fan2023dataset}, the fatigue database of complex metallic alloys consists of 272 entries~\cite{zhang2023fatigue}, while other mechanical properties of high-entropy alloys tabulations contain $\sim$370 entries~\cite{gorsse2018database}.
These limitations hinder the application of ML in alloys. In our workflow, we employ textual inputs generated by LLMs, offering the advantage of eliminating the need for manual feature extraction. The second tier in our workflow is the application of LLMs to output content relevant to specific problems in materials science with customized prompts.
As LLMs possess considerably more parameters than traditional ML models, they generate a richer and more diversified knowledge base.

In the third tier, a  BERT model, employing an attention mechanism for Self-Instruct learning approach for automatic feature extraction, provides classification.
Different pre-trained models are utilized for different downstream tasks.
Models based on the attention mechanism are not limited to text input but may also be applied to areas like image recognition~\cite{Li2023} and molecular structure identification~\cite{zhou2023uni}.
These features allow our workflow to seamlessly support a combination of different datasets and multimodal expansion, addressing the issue of insufficient datasets for specific material problems often encountered in traditional methods.

Finally, the material classification models trained through our workflow can be applied to numerous downstream tasks, such as discovering new materials for different technologies, conducting interpretability analysis to extract composition-structure-property relationships, and even fine-tuning models to achieve even better results in other classification tasks.
Our workflow represents a novel, general approach for material sciences by constructing large-scale databases from prompt-engineered LLMs and training general DL models to expedite materials discovery.
We showcase the efficiency and accuracy of the workflow by applying it to the MG classification problem described above.

\section*{Application: metallic glass classification}
MGs are amorphous structures typically containing several metals and non-metallic elements~\cite{VARSHNEYA20191}.
The family of MGs is renowned for superior engineering performance, including high strength, high thermal stability, and excellent corrosion resistance~\cite{halim2021metallic}.
Despite ongoing efforts to discover MGs with enhanced comprehensive performance, their formability remains a complex and challenging problem in materials science.
Predicting glass formability (thermal quench rates required for glass formation) is also difficult for atomistic simulations due to large sample size and, especially, the long timescale (inaccessible to first-principles calculations) and the absence of accurate interatomic potentials for molecular dynamics simulations.
Moreover, the compositional space of MGs is of high dimension, often comprising more than three elements, rendering the experimental search for new applicable MGs a time-consuming process~\cite{liu2023effective, li2017many}.

\begin{figure*}[t]
  \centering
  \includegraphics[width=0.95\textwidth]{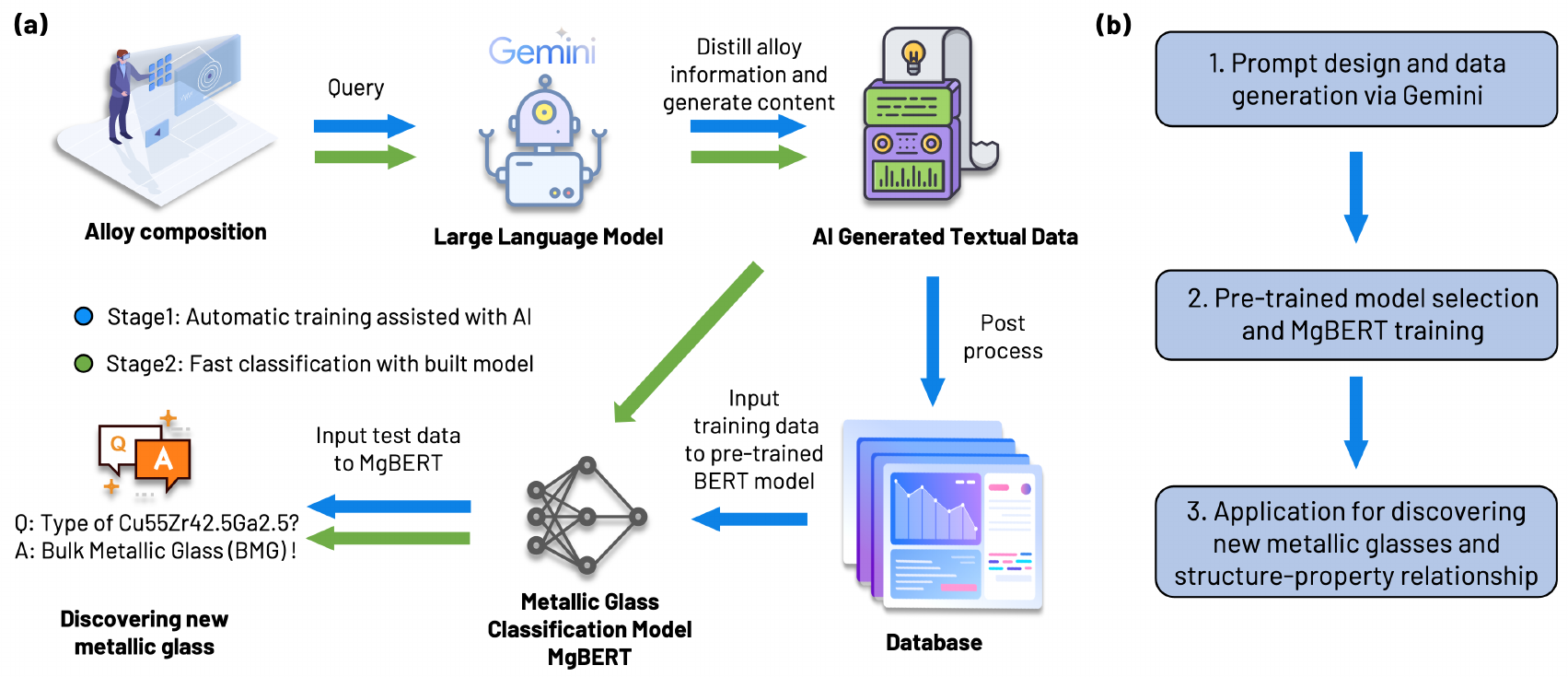}
  \caption{Application of our workflow for the classification of metallic glasses. (a) Schematic for the customized workflow, from known alloy composition to the discovery of new metallic glasses. (b) Three essential steps in the customized workflow include converting alloy composition to textual data, training classification models, and discovering and interpreting new metallic glasses.}\label{fig2}
\end{figure*}

Various ML methods~\cite{zhou2022critical} have been employed to classify MGs, usually based upon manually calculated or carefully designed input material features.
The inconsistency of designed features and datasets across different ML models often necessitates recalculating those features when utilizing other models.
In the worst case, experiments must be rerun to provide input to different models.
In this regard, we leveraged existing MG experimental data~\cite{ward2018machine} and applied it to the proposed workflow here.
The dataset compiled by~\cite{ward2018machine} consists of 8,415 alloy compositions classified according to MG formability.
The alloy compositions were classified into three categories based on the critical cooling rate ($R_{\text c}$), BMG ($R_{\text c} < 10^3 \text{ K/s}$), ribbon ($10^3 \text{ K/s} < R_{\text c} < 10^6 \text{ K/s}$), and non-ribbon (NR, $R_{\text c} > 10^6 \text{ K/s}$); i.e., the glass formability is then BMG $>$ ribbon $>$ NR.
From this dataset, we focus on 5,577 alloy compositions, of which 80\% were used in the training set - the remainder forming the test set.
As shown in Fig.~\ref{fig2}a, a customized version of the general workflow was developed for this input data type.
The blue arrows indicate the training process, which involves expanding the information on each alloy (textual data - e.g., physical properties) using an LLM.
Subsequently, a DL model is trained on the textual data from the LLM-expanded training set for MG classification.
Finally, the efficiency and accuracy are analyzed using the textual data from the test set.
We then query the model to interpret the results.
The green arrows in Fig.~\ref{fig2}(a) indicate the application stages of the model; the three main steps (below) are illustrated in Fig.~\ref{fig2}(b).
The workflow can determine the MG category (BMG, ribbon, or NR) within seconds for an unexplored alloy composition.

\begin{figure*}[t]
  \centering
  \includegraphics[width=0.95\textwidth]{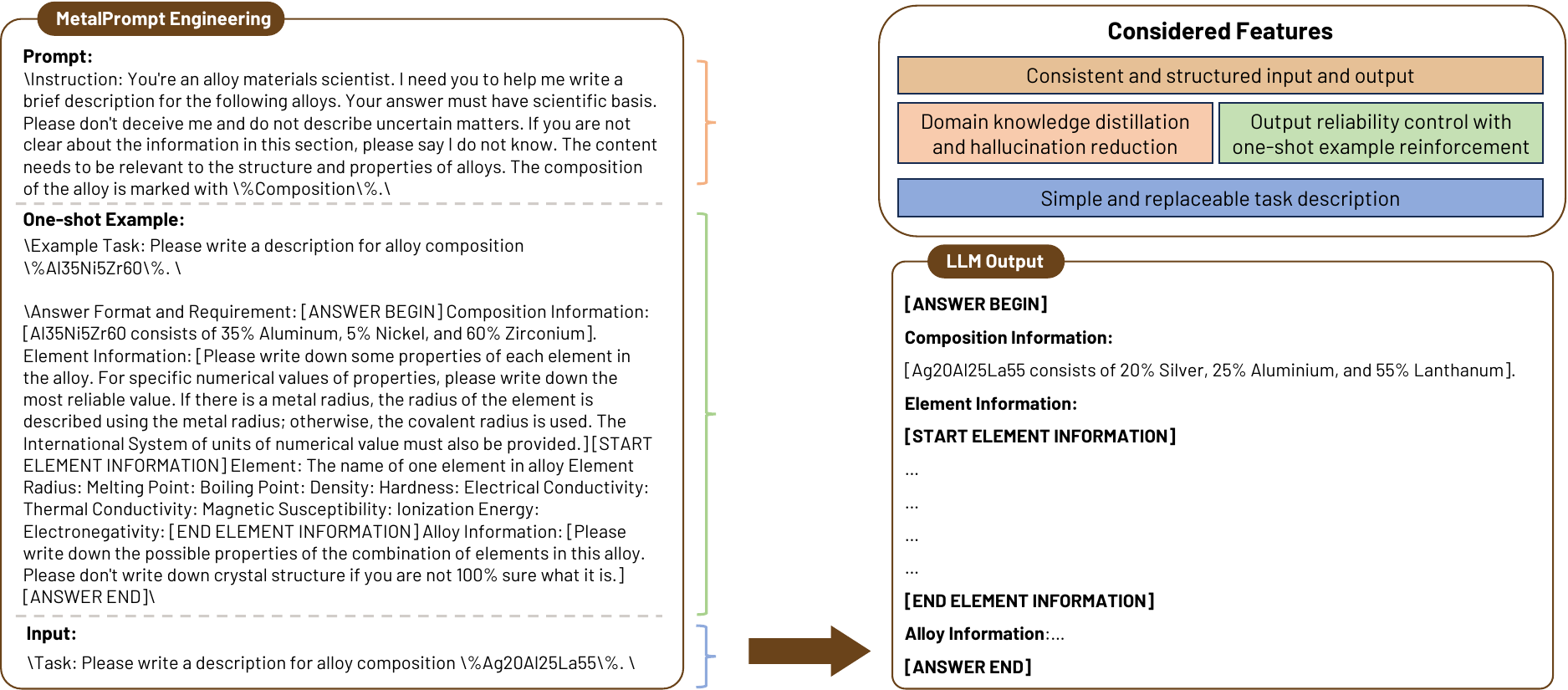}
  \caption{Schematic of ``MetalPrompt'' for generating textual data with prompt engineering. The left part depicts the schematic text for the input query of the LLM. The sections enclosed within brackets of different colors (on the right) represent diverse features considered in prompt engineering.}\label{fig3}
\end{figure*}

In the first step, we drew inspiration from a previous design of metal-organic framework prompt engineering~\cite{zheng2023chatgpt} and designed a prompt, ``MetalPrompt'', to generate textual descriptions for various alloy compositions.
Fig.~\ref{fig3} illustrates the ``MetalPrompt'' design concept.
The schematic text for the input query of the LLM is on the left, consisting of a prompt section, a one-shot example section, and an input section.
The prompt section enables the LLM to focus on generating domain-specific knowledge while minimizing hallucinations.
The one-shot example section ensures a consistent output structure and incorporates the information of interest.
The output focuses on alloy composition (atomic percent of each element), elements (thermodynamic properties of each element), and alloy physical/chemical properties.
The input section ensures that the task description is simple and interchangeable, thus preventing disruption of the model output by irrelevant information.
We have added some emphasis symbols such as ``\textbackslash \textbackslash'' at the beginning and end of the paragraph, as well as ``\textbackslash\%composition\textbackslash\%'' symbols for compositions, guiding the model to recognize critical information and structural layering in the input~\cite{rodriguez2023prompts}.
Supplementary Figs.~S1-S3 show the output without ``MetalPrompt'' in different LLMs; the output text structure is inconsistent and is subject to hallucinations.
Supplementary Figs.~S4 and S5 compare the prompt baseline and ML models, and show the effects of using different prompt methods.
In both comparisons, our ``MetalPrompt'' demonstrates the best performance.

\begin{figure*}[t]
  \centering
  \includegraphics[width=0.95\textwidth]{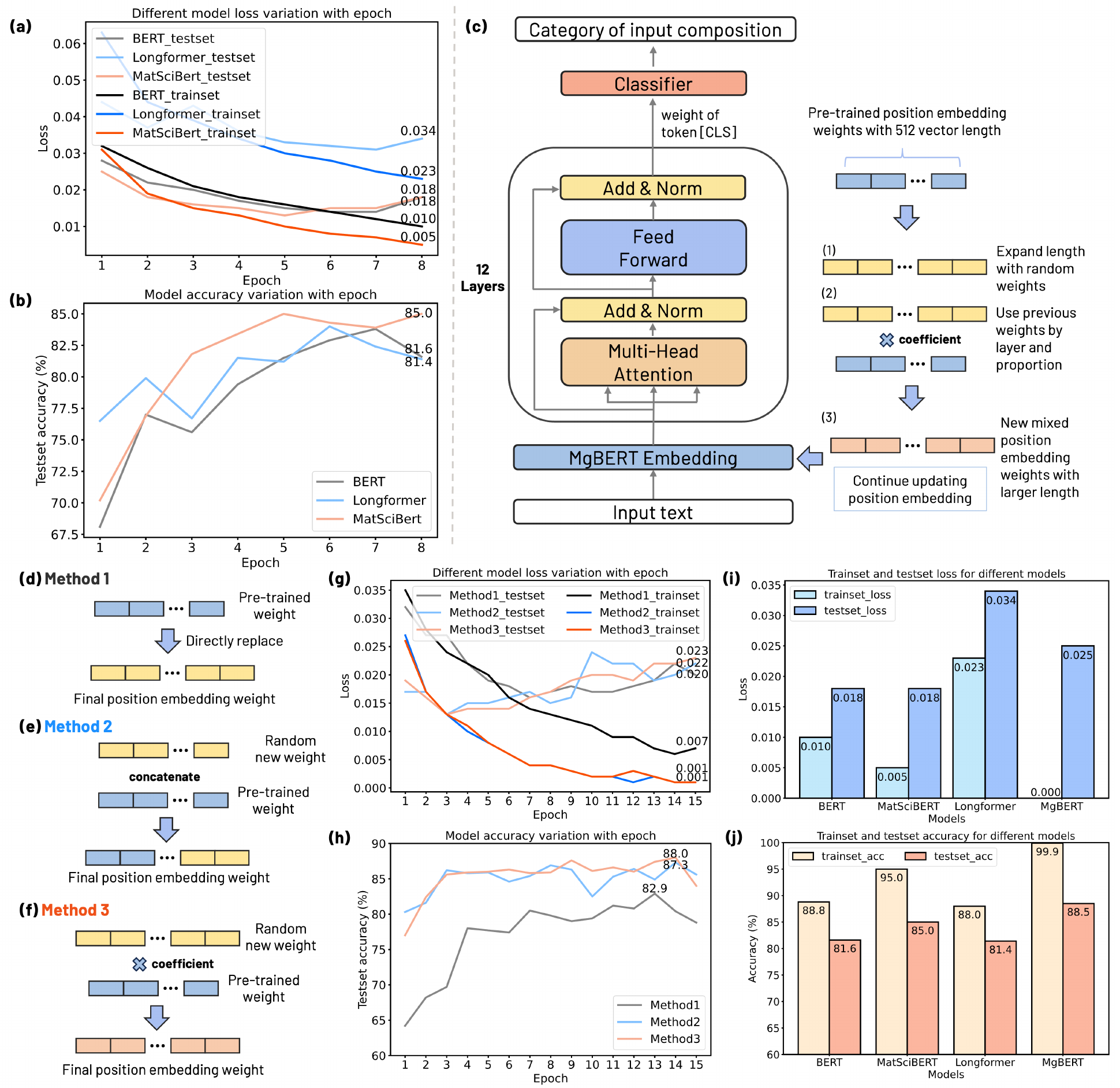}
  \caption{Performance comparisons and architecture of MgBERT. (a, b) performance comparison of various pre-trained models. (c) fundamental architecture of the MgBERT model and schematic diagram of the working principle of MgBERT position embedding. (d, e, f) three distinct position embedding methods. (g, h) performance comparison of different embedding techniques. (i, j) performance comparison between MgBERT and other pre-trained models.}\label{fig4}
\end{figure*}

\begin{figure*}[t]
  \centering
  \includegraphics[width=0.95\textwidth, trim={0 0.1cm 0 0}, clip]{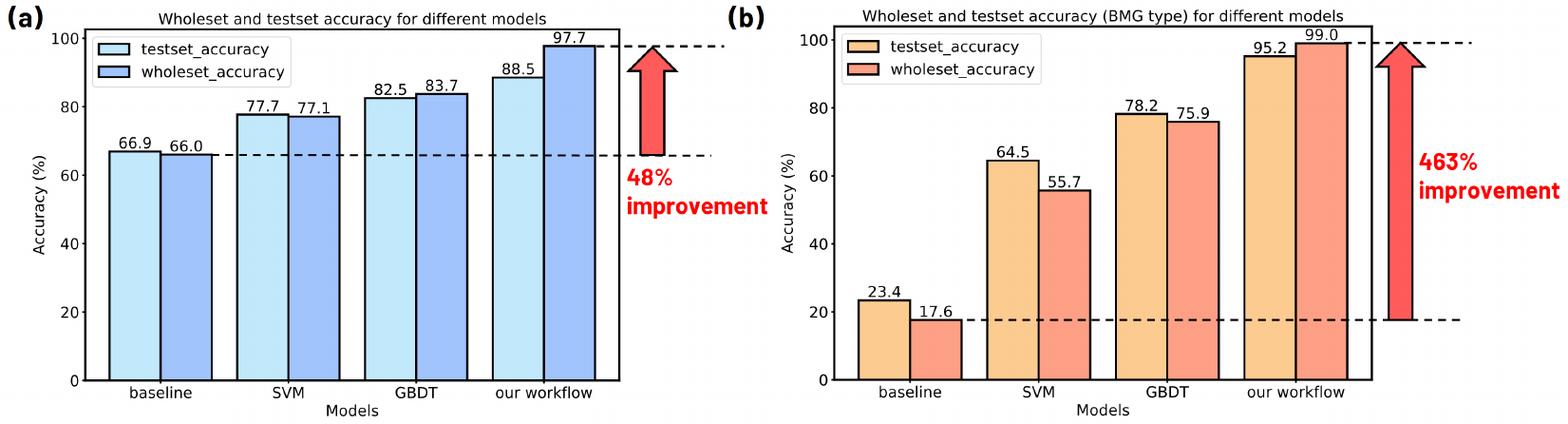}
  \caption{Comparison of classification accuracy between our workflow and traditional machine learning models. (a) comparison in classifying all category tasks. (b) comparison in classifying bulk metallic glasses.}\label{fig5}
\end{figure*}

Next, we trained an MG BERT (MgBERT) classification model combining Self-Instruct learning and supervised classification.
DL models for natural language processing typically require a vast number of fitting parameters, necessitating the selection of a suitable pre-trained model as a base. MgBERT is fine-tuned based on the MatSciBERT (a pre-trained model based on materials science texts)~\cite{gupta2022matscibert}. Figs.~\ref{fig4}(a, b) compare different pre-trained BERT models, with that based upon MatSciBERT achieving the best classification accuracy and model loss on the test set.
While MatSciBERT only accepts inputs of up to 512 tokens, our text data has a maximum length of 859 when converted to tokens.
Consequently, we designed a dynamically resizable embedding layer called ``MgBERT Embedding'' to meet the input scale requirements.

This embedding layer assigns weights based on layers and proportions according to the preset length of the new token, building upon the embedding weights trained by MatSciBERT.
This approach generates new position embeddings that support longer text inputs and circumvent time-consuming weight initialization training.
This position embedding layer (in Fig.~\ref{fig4}(c)) is employed to tokenize the input text of the model.
Comparison of the results in Figs.~\ref{fig4}(g) and (h) show that the MgBERT embedding Method 3 (Fig.~\ref{fig4}(f)) exhibits higher accuracy and lower training set loss, compared to the embedding layer method in Fig.~\ref{fig4}(d) (using newly initialized weights), and directly using pre-trained weights (covering the first 512 embedded positions - see Fig.~\ref{fig4}(e)).
For different pre-trained models based on the same text data (Figs.~\ref{fig4}(i, j)), our MgBERT has the best performance in terms of the training set loss, training set classification accuracy, and test set classification accuracy.
Notably, for alloy data in the test set (not used in the training), our model achieved an accuracy of 88.5\% (an improvement of 9\% compared to Longformer and an improvement of 4\% compared to the best performing MatSciBERT).
The detailed classification results are shown through a confusion matrix in Supplementary Fig.~S6.
These findings showcase the success of our improvement in embedding structure, simultaneously supporting longer input text and enhanced accuracy.

To examine the effectiveness of the entire workflow, we trained a baseline model based on logistic regression (LR) using the training set, as well as a support vector machine (SVM) model and a gradient boosting decision tree (GBDT) model.
We tested the results (on the test set) and compared the classification performance of different methods (Fig.~\ref{fig5}).
For the default three-category MG classification problem, our workflow achieved a maximum accuracy increase of 48\% across the entire dataset and a peak enhancement of 32\% on the test set (as compared with the baseline).
The accuracy of the workflow in identifying BMGs, as shown in Fig.~\ref{fig5}b, increased by 463\% on the entire dataset and by as much as 307\% on the test set as compared with the baseline (traditional ML) classification model.
Considering that BMGs constitute only $\sim$11\% of our dataset, these results highlight the capability of our text-based material classification workflow in accurately recognizing patterns in data-scarce scenarios, far surpassing traditional ML models.

\begin{figure*}[!ht]
\centering
\includegraphics[width=0.93\textwidth]{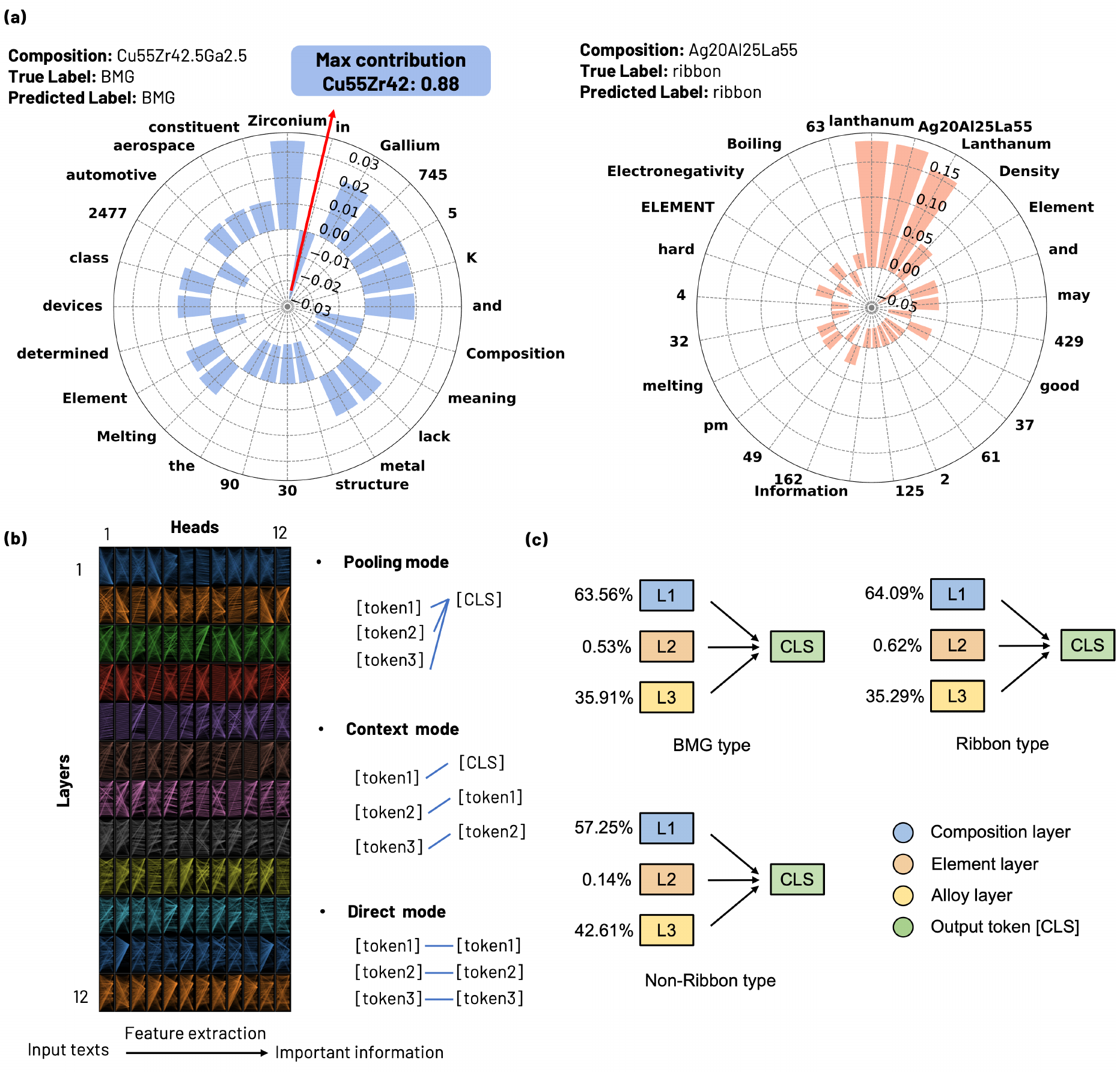}
\caption{Interpretability for classification deep learning model. (a) contribution of input text to the classification results, with the outward (inward)-facing bar on the polar axis representing a positive (negative) contribution. (b, c) Attention-level analysis. (b) attention score flow diagram across different layers and heads. The model is a 12-layer deep neural network where each layer has 12 heads (each head is a neural network). Input/output for each head consists of 768 tokens (representations of individual words in the text input). Different colors represent different layers, while lines represent connections between the input and output tokens of each head. Line thickness indicates the attention score: a larger attention score implies that the output token focuses more on a particular position in the input. Three modes on the right represent typical patterns of output token information extraction from input tokens. The pooling mode demonstrates input information aggregation to the same output token, the context mode shows input aggregation to the output token at ``position-1'' (token 2 to 1), and the direct mode indicates direct input transfer to the output token at the same position. The output from 12 heads in each layer is concatenated and passed to the next layer, enabling the extraction and transfer of crucial input text information to the final classifier. (c) attention pooling diagram of the final attention layer of MgBERT. To facilitate differentiation, the input text is divided into three text layers. Blue: composition layer consisting of tokens before ``element information'' in the input text (see Fig.~\ref{fig3}), Orange: element layer comprising tokens before ``alloy information'', and Yellow: alloy layer containing the remaining tokens. The number on the left indicates the last attention layer [CLS] (classify) token attention to various text layers.}\label{fig6}
\end{figure*}

DL models often lack algorithmic transparency~\cite{chakraborty2017interpretability}.
Language models such as BERT typically have hundreds of millions of parameters~\cite{devlin2018bert}, making it challenging to explain the model results (at the algorithmic level).
We applied the local interpretable model-agnostic interpretations (\emph{LIME})~\cite{ribeiro2016should} method to analyze the input-output relationship.
We selected text data of alloy compositions Cu$_{55}$Zr$_{42.5}$Ga$_{2.5}$ (BMG), Ag$_{20}$Al$_{25}$La$_{55}$ (ribbon), and Al$_{40}$Mn$_{25}$Si$_{35}$ (NR) for this analysis. Fig.~\ref{fig6}(a) and Supplementary Fig.~S7 show to what degree the words in the input data from the LLM (marked LLM Output in Fig.~\ref{fig3}) are most important for the input/output correlation.
Fig.~\ref{fig6}(a) shows that the maximum contribution to this correlation is the composition Cu55Zr42, with a value of +0.88.
On the other hand, for the Ag20Al25La55 case, the alloy contribution to the correlation is much weaker.
Since the system ``knows'' the atomic sizes and that those atomic sizes are much different in the BMG CuZr alloy cases (compared with the ribbon AgAlLa case), this may suggest that atomic size difference is important, consistent with the classical Inoue rule~\cite{inoue2000stabilization}  for glass formability and high-throughput experiment results~\cite{ren2018accelerated} (i.e., easy glass formation for more than three atom types with atomic size ratios above 12\%).
The model also shows that some properties of individual elements are important.
For example, ``Melting'' refers to elemental melting points, which is also used as one of the parameters commonly employed as a predictor for glass formation~\cite{johnson2016quantifying}.
The word ``745'' in the CuZr alloy in Fig.~\ref{fig6}(a) refers to the ionization energy of copper - this may influence the tendency for charge transfer between Cu and other alloying elements.
The word ``2477'' represents the boiling point of gallium; elemental boiling points have been suggested as related to some MG behavior~\cite{wang2009bulk}.

These findings suggest that our text classification approach can be used to automatically extract features from the textual input.
Building on this foundation, in contrast to traditional ML models that rely upon manually designed features as input, our text-based workflow enables models to accept any raw text as input.
This flexibility allows for rapidly transferring and recycling pre-trained model parameters between different materials classification tasks.
We also visualized the attention scores within the classification model, and the model extracts features based on three distinct patterns (Fig.~\ref{fig6}(b)).
An attention score tells the importance of the contribution of each ``neuron'' strength in this 12-layer deep neural net.
This process ensures that after 12 layers of training, the output [CLS] (classify) token possesses sufficient information for classification.
Since we use the value of the [CLS] token from the final layer to make judgments in the classification layer, Fig.~\ref{fig6}(c) visualizes the importance of different sections of the input text.
The results suggest that the composition and alloy layers have garnered significant attention (compared with the element layer - element information is implicitly contained in the other two layers).
A possible explanation is that key tokens within these two layers have captured the essential features from the input text.
Consequently, the [CLS] token can achieve accurate classification results by focusing on those specific token positions.
These findings demonstrate that the classification MgBERT model, generated by our general workflow, has the abilities of dynamic feature extraction, modeling inter-word relationships, hierarchical representation learning, and bidirectional context understanding, which are indicators of good classification performance.

\section*{Discussion}
We emphasize that classification cannot be successfully achieved using direct questioning alone (as seen in Supplementary Figs.~S8-S10).
Therefore, as part of our workflow, we designed a BERT-based classification model to categorize materials using text generated by the LLM.
In practice, employing a classification model is a necessary step now (future, more materials robust, LLMs may not need this step), as directly querying the LLM yielded a near-zero accuracy rate.

The general workflow proposed here offers a novel approach to address a significant challenge in ``AI for material science'', where datasets are often small.
By utilizing a pre-trained BERT model (e.g., MatSciBERT optimized for material science), we can fine-tune the pre-trained model with textual data from a prompt-engineered LLM to achieve higher accuracy classification  (up to 99.0\% for the entire dataset).

We propose a general approach for materials classification and demonstrate its efficiency for a specific example, i.e., the classification of MGs.
Our general workflow can be applied to any material applications, provided that labeled material samples are available.
The findings point toward a future where integrating advanced language models and domain-specific fine-tuning will revolutionize material classification, especially in the very common cases where the data is sparse.
We envision that LLMs may eventually serve as ``world models''~\cite{dawid2023introduction}, encapsulating knowledge across diverse materials and domains.
Consequently, such models based on text for classification, prediction, or even generating new materials will become potential tools for addressing challenges in material selection/design.
A sufficiently robust LLM can provide desired classification outcomes directly through appropriate prompting and significantly streamline the process of materials discovery and design.


\section*{Methods}
\subsection*{Data processing for metallic glasses dataset}
We curated the original dataset for metallic glasses~\cite{ward2018machine} by deleting $\sim1/3$ of the entries (total 8,415) for which the original classification was ambiguous.
The result was a dataset of 5,577 alloys and their respective classifications.
To implement prompt engineering, we employed the \emph{langchain} library~\cite{langchain} to systematically apply the template shown in Fig.~\ref{fig3} to each alloy in the dataset.
We divided the 5,577 data points into a training and test set at a ratio of 80:20 to train the MgBERT model.
The accuracy of the MgBERT model was assessed using the test set and compared with traditional machine learning models.
We employed the \emph{Matminer}~\cite{WARD201860} and \emph{Pymatgen}~\cite{ONG2013314} tools to convert the chemical formulas of alloys into ``composition features'' as defined within the \emph{Matminer}~\cite{WARD201860} framework.

\subsection*{LLM and prompt engineering}
In this work, we employ the Google \emph{Gemini-pro}~\cite{geminiteam2023gemini} LLM, which surpasses human experts in massive multi-task language understanding~\cite{hendrycks2021measuring}.
As discussed earlier, our research harnesses the automated orchestration capabilities of the \emph{langchain}~\cite{langchain} library to script the bulk generation of prompts and textual data.
This scripting forms an integral part of a generic workflow and offers high adaptability.
The prompt templates developed here are not rigid constructs but are designed to be inherently flexible, allowing for arbitrary modifications tailored to researcher demands and objectives.
This level of customization is critical in materials science, where the nuances of the subject matter vary significantly from study to study.
The ability to fine-tune prompts to align with particular research questions or datasets ensures that the language model can be effectively leveraged to generate practical and contextually relevant textual data outputs.

Our investigation sought to demonstrate the effectiveness of prompt engineering by creating a range of alternative prompt templates to act as comparative benchmarks.
These templates comprised one that integrated a chain-of-thoughts (CoT) approach with a few-shot learning paradigm (denoted ``few-shot with CoT''), another that depended exclusively on few-shot learning examples (denoted ``few-shot''), a third that was based on a zero-shot learning scenario (denoted ``direct inquiry''), and our custom-designed ``MetalPrompt'' from our established workflow.
By leveraging these diverse templates, we crafted specific prompts that were subsequently employed to interrogate the \emph{Gemini-pro} LLM.
The underlying rationale for this prompt strategy was to gather and juxtapose accuracy metrics for different prompting methodologies.

\subsection*{MgBERT training and evaluation}
Starting from MatSciBERT, we implemented a classifier function with MgBERT embedding to build MgBERT.
Accuracy and cross-entropy are used to evaluate the model's effectiveness. 
For classification within MgBERT, we used the value of the [CLS] token in the last layer of the attention section of MgBERT as the output to aggregate information for the entire model.
Then, through a classifier, we compress the value of the token into a vector of length 3, and label 0 as BMG, 1 as Ribbon, and 2 as Non-ribbon.
We then identified the maximum position of the output vector value and used this as the final classification result.
The overall expenses associated with MgBERT training was $\sim$\$46 USD on an Nvidia V100 (i.e., $\sim$30 hours at $\sim$\$1.53 USD/hour).
The cost of 1.89 million input and 2.14 million out tokens was $\sim$\$1.5 USD (official \emph{Gemini-pro} pricing on January 29, 2024).
To apply MgBERT for inference on 1,000 different compositions cost  $\sim$\$0.3 USD.
This training and application of MgBERT enables researchers to cost-effectively apply this general approach across a wide range of materials applications.

\subsection*{Interpretability method}
To elucidate input and output relationships, we applied the \emph{LIME}~\cite{ribeiro2016should} (local interpretable model agnostic explanations) technique; widely used to explain predictions of machine learning models by constructing relationship functions via interpolation sampling.
Here, we displayed the 25 most important features and performed 1,000 interpolation samples for function fitting.

To analyze the input-output relationship at the attention level, we used the \emph{bertviz}~\cite{vig-2019-multiscale} library to complete the attention score flow diagram for 12 layers and 12 attention heads.
To attribute the importance of the [CLS] token in the final attention layer, we first averaged the attention score of the 12 attention heads in the last layer, then divided the input text words into three layers based on their positions (i.e., composition, element, and alloy).
To evaluate the importance of each layer, we calculated the sum of attention scores within that layer, weighted by the token length ratio, and then divided by the total token length.

\section*{Data availability}
The method section provides the models and algorithms employed in this study, while specific parameter implementations can be found in the Supplementary Information (SI).

\section*{Code availability}
All the codes used in the paper will be made available on GitHub upon the acceptance of manuscript, and these codes will be provided upon reasonable requests.

\def\bibsection{\section*{\refname}}
\bibliography{./sn-bibliography.bib}

\section*{Acknowledgments}
This work is supported by Research Grants Council, Hong Kong SAR through the Collaborative Research Fund (C1005-19G) and General Research Fund (17210723). T.W. acknowledges additional support by The University of Hong Kong (HKU) via seed funds (2201100392, 2309100163). S.P. acknowledges additional support by HKU via seed fund (2309100201). We acknowledge helpful discussions with Prof. Yi Ma at HKU and Simon Zhai at UC Berkeley.

\section*{Competing interests}
The authors declare no competing interests.

\section*{Supplementary Information}
Supplementary Notes

Supplementary Figures S1-15

Supplementary Tables S1-2

References (56-85)

\clearpage
\pagebreak
\beginsupplement
\pagestyle{plain}
\onecolumngrid

\begin{center}
  \huge {Supplementary Information for}
  \bigskip

  \large \textbf{A Prompt-Engineered Large Language Model, Deep Learning Workflow}

  \large \textbf{for Materials Classification}
  \bigskip
  \bigskip

  Siyu Liu, Tongqi Wen, A. S. L. Subrahmanyam Pattamatta, David J. Srolovitz
\end{center}

\clearpage

\makeatletter

\titleformat{\subsubsection}
{\large}{\thesubsubsection}{1em}{{#1}}

\titleformat{\subsection}
{\bfseries\large}{\thesubsection}{1em}{{#1}}

\titleformat{\section}
{\bfseries\Large}{\thesection}{1em}{{#1}}

\section{Supplementary Notes}
\subsection{Glossary}
\begin{itemize}
\item \textbf{Artificial Intelligence}: artificial intelligence (AI) refers to the development of computer systems capable of performing tasks that typically require human intelligence, including visual perception, decision-making, and language comprehension~\cite{JOINER20181}.
\item \textbf{Attention}: a mechanism enabling models to focus on specific parts of the input sequence during processing, allowing them to weigh the significance of different words when generating output. The attention mechanism assists in capturing long-range dependencies and enhancing performance in various natural language processing (NLP) tasks~\cite{NIU202148}.
\item \textbf{BERT}: Bidirectional Encoder Representations from Transformers, a widely used pre-trained language model~\cite{devlin2018bert}.
\item \textbf{Chain-of-thoughts}: a prompt engineering method that enhances complex reasoning capabilities by adding intermediate reasoning steps to the text input for the model~\cite{wei2023chainofthought}.
\item \textbf{ChatGPT}: Chat Generative Pre-trained Transformer, developed by OpenAI, is a chatbot and a sibling model to InstructGPT, trained to follow instructions in a prompt and deliver detailed responses~\cite{RAY2023121}.
\item \textbf{Cross Entropy Loss}: a performance indicator measuring classification models, also known as log loss, with lower values indicating better predictive performance.
\item \textbf{Deep Learning}: deep learning (DL) is an AI technology emulating the neural network structure of the human brain, learning and training via multi-level neurons for complex data analysis and processing~\cite{helm2020machine}.
\item \textbf{Embedding}: an NLP technique that maps a high-dimensional space, with a dimensionality equal to the total number of words, to a much lower-dimensional continuous vector space, assigning each word or phrase a vector in the real number domain~\cite{almeida2023word}.
\item \textbf{Feature Engineering}: selecting, transforming, and creating new features from raw data to improve machine learning (ML) model performance. It involves identifying relevant and informative input variables and preparing them for accurate predictions or classifications~\cite{zheng2018feature}.
\item \textbf{Feed Forward}: refers to a deep learning framework, also known as a multi-layer perceptron (MLP), that defines a mapping $y=f(x:\theta )$ and learns the value of parameter $\theta$ to achieve the best function approximation~\cite{svozil1997introduction}.
\item \textbf{Few-shot Learning}: an ML approach where models are trained to make accurate predictions with limited examples per class~\cite{10.1145/3386252}.
\item \textbf{Gemini}: Google's multimodal LLM, the first to outperform human experts on MMLU (Massive Multitask Language Understanding), a popular method for testing the  knowledge and problem-solving abilities of AI models~\cite{geminiteam2023gemini}.
\item \textbf{Gradient Boosting Decision Tree (GBDT)}: also known as Gradient Boosting Machines (GBM), is a popular ensemble learning method combining decision trees with gradient boosting. GBDT continuously fits new models for more accurate estimates of response variables. The principle behind this algorithm is to construct new base learners that maximize correlation with the negative gradient of the ensemble's loss function~\cite{natekin2013gradient}.
\item \textbf{Hyperparameters}: external configuration variables input in advance to manage the training of ML models~\cite{yang2020hyperparameter}.
\item \textbf{In-context Learning}: refers to enhancing LLM performance using a few examples in the input context~\cite{dong2023survey}.
\item \textbf{ImageNet}: an image database organized according to the WordNet hierarchy (currently only nouns), with each hierarchy node represented by hundreds of thousands of images~\cite{deng2009imagenet}.
\item \textbf{Knowledge Distillation}: refers to extracting knowledge from larger deep neural networks into smaller networks~\cite{gou2021knowledge}.
\item \textbf{Large Language Model}: large language model (LLM) is a class of DL models trained on large amounts of text data to learn statistical patterns of natural language~\cite{brown2020language}.
\item \textbf{LLM Hallucination}: occurs when LLMs generate content deviating from user input, contradicting previously generated context, or inconsistent with established world knowledge~\cite{zhang2023sirens}.
\item \textbf{Logistic Regression}: a supervised learning algorithm utilizing logistic functions to estimate label probabilities~\cite{menard2002applied}.
\item \textbf{Machine Learning}: machine learning (ML) demonstrates the experiential `learning' associated with human intelligence, along with the ability to enhance its analyses through using computational algorithms~\cite{helm2020machine}.
\item \textbf{Multi-head Attention}: represents multiple attention modules within a single attention layer of the model, allowing for different focus on various parts of a sequence.
\item \textbf{Pre-trained Model}: a model trained on a large corpus of data that may be fine-tuned to solve various tasks~\cite{han2021pre}.
\item \textbf{Prompt Engineering}: a method to generate textual data from LLMs by embedding task descriptions in the input, effectively conveying specific parameters to the model as part of a problem statement~\cite{RAY2023121}.
\item \textbf{Self-Instruct Learning}: a method for improving the instruction-following capabilities of pre-trained language models by bootstrapping off their generations~\cite{wang2023selfinstruct}.
\item \textbf{Supervised Classification}: an ML task that categorizes input data into predefined classes or categories based on labeled training examples~\cite{antonelli2019integrating}.
\item \textbf{Support Vector Machine}: a supervised learning algorithm for classification and regression tasks that identifies the optimal boundary (hyperplane) separating data points of different classes~\cite{suthaharan2016support}.
\item \textbf{Tokenization}: the process of converting text into smaller structural markers, called tokens. The tool used to handle this process is known as a tokenizer~\cite{kudo2018sentencepiece}.
\item \textbf{Transfer Learning}: an ML technique employing a pre-trained model as the starting point for a new related task instead of training a model from scratch. It allows for different domains, tasks, and distributions in training and testing~\cite{pan2009survey}.
\end{itemize}

\subsection{Workflow details}
\subsubsection{Data processing of metallic glasses dataset}
As mentioned in the ``Methods'' section, our initial dataset consists of 5,577 samples.
To train different BERT models, it is necessary to divide it into a training and a test set.
We employed a stratified sampling method for this purpose, ensuring a representative and balanced representation of the different subgroups~\cite{TAL2011229}, which enhances the quality and reliability of the sampling.
We  performed non-repetitive stratified sampling on three alloy data categories based on the overall 80:20 split ratio for the training and test sets.
Ultimately, we obtained a training set with 4,460 samples and a test set with 1,117 samples, maintaining the 80:20 ratio for each category.

The textual data generated through \emph{Gemini-pro} is used to train different BERT classification models.
The data format of obtained data is shown in Table~\ref{tableS1}.

\subsubsection{Prompt design and textual data generation}
Upon acquiring the alloy composition data, we proceeded with prompt design.
Prompt engineering is a prominent direction in the field of LLMs and is considered a method that significantly improves data generation effectiveness for LLMs~\cite{liu2023pre}.
Techniques such as few-shot learning~\cite{wei2023chainofthought}, CoT~\cite{wei2023chainofthought}, synergizing reasoning and acting (ReAct)~\cite{yao2023react}, and retrieval-augmented generation (RAG)~\cite{lewis2021retrievalaugmented} are  effective prompt engineering methods.
Due to the additional knowledge sources required for the latter two methods, we tested direct inquiry, few-shot, and CoT, and compared them with our MetalPrompt.
Figs.~\ref{figS8}, \ref{figS9}, and \ref{figS10} are templates for three benchmark methods, while Fig.~\ref{figS4} shows the comparison results.
Our ``MetalPrompt'' achieved the best performance.

For the prompt inquiry setting with the LLM, we used Top-K = 1 and Temperature = 0 as default parameters.
As shown in Figs.~\ref{figS11} and \ref{figS12}, a smaller Top-K and lower Temperature indicate more reliable model output.
Top-K = 1 implies that the model performs greedy decoding and selects the most probable value.
Unless specifically explained, the language model parameters mentioned below are consistent with these settings.

Using the above parameter settings and ``MetalPrompt'' as the input template, we generated textual data for 5,577 alloy compositions.
The input of alloy elements in the template was assisted by the \emph{langchain}~\cite{langchain} library.
Simultaneously, based on the alloy composition category relationship of the original training and test dataset, we labeled the generated textual data with the same category.
This dataset is used for training the classification model below, partitioning training, and testing sets consistent with the original data.

\subsubsection{MgBERT training and evaluation}
\textit{\textbf{BERT model and attention mechanism:}}
BERT~\cite{devlin2018bert} is a pre-training language representation method that has had significantly impact in the  NLP field.
Traditional language models analyzed text data unidirectionally, either from left to right or right to left, limiting their understanding of language context.
Using the Transformer architecture, BERT processes each word concerning all other words in a sentence rather than sequentially. 
BERT was selected as the foundation for our classification model due to its state-of-the-art results in numerous tasks.

Fig.~\ref{fig4}c in the main text shows the basic BERT architecture, comprising data pre-processing, input encoding, model training, post-processing, and output results.
During data processing, tokenizers divide sentences or words into individual tokens, such as replacing ``for example'' with ``for'' and ``example''.
Then, as demonstrated in Fig.~\ref{figS13}, the encoding method transforms pre-processed text data into the input representation of the model.
For instance, ``for'' is converted to a vector $[00..1..00]$. 
Model training is crucial to BERT and we use multi-head attention to extract essential features from input embeddings.
Attention mechanisms enhance model performance in NLP and other sequence data processing tasks by focusing on the most relevant parts of the input sequence.
Models can learn to assign various attention weights to input information from different positions, concentrating on crucial aspects when processing input sequences.
This mechanism allows the model to capture long-distance dependencies and important patterns in sequences more effectively, thus improving its performance in processing sequence data~\cite{vaswani2023attention}.
Fig.~\ref{figS14} shows a schematic diagram of the operation of the attention mechanism.
Compared to a single module, multi-head attention employs multiple attention modules in the same calculation, extracting different feature information.
The calculation formula for attention score is as follows:
\begin{equation}
	\textrm{Attention}(Q, K, V) = \textrm{softmax}(\frac{QK^T}{\sqrt{d_k}})V
\end{equation}
\begin{equation}
	\textrm{MultiHead}(Q, K, V) = \textrm{Concat}({\textrm{head}_1, ..., \textrm{head}_\textrm {h}})W^O,
\end{equation}
where $\textrm{head}_\textrm{i} = \textrm{Attention}(QW{i}^{Q}, KW{i}^{K}, VW{i}^{V})$, $\sqrt{d_k}$ refers to the queries and dimension keys, $\textrm{softmax}$ refers to the softmax function, and  $Q$, $K$, and $V$ refer to the input query, key, and value, respectively.

\textit{\textbf{Pre-trained models selection:}}
We compared three pre-trained models based on BERT, including the basic BERT model~\cite{devlin2018bert}, Longformer supporting longer input text lengths~\cite{beltagy2020longformer}, and MatSciBERT trained on materials science texts~\cite{gupta2022matscibert}.
All models employ default parameter settings; however, to expedite calculations, the maximum input length for Longformer is limited to 1200 tokens, while other models have a default maximum length of 512.
To ensure reproducibility, the random seed is fixed at 42. 
All models are trained using cross-entropy loss and optimized with the Adam optimizer for parameters.
All training was conducted on a 32G Nvidia V100. 
Table~\ref{tableS2} presents the basic hyperparameters for model training.
Due to its larger embedding layer requiring more graphics memory, Longformer has a slightly smaller batch size.

\textit{\textbf{MgBERT embedding and training:}}
In training of MgBERT we found that the file with the longest tokens is ``Fe68.3C6.9Si2.5B6.7P8.8Cr2.2Mo2.5Al2.1.txt'' with 859 tokens, and 300 files have more than 512 tokens.
Our basic model, MatSciBERT, only supports inputs with a maximum token length of 512.
Considering the potential inconsistency in text input length for different materials and scalability requirements, we designed a variable-length MgBERT position embedding layer that does not necessitate complete retraining.
The following Algorithm 1 is a pseudocode representation of its principle.

To compare three embedding methods, we ran 16 epochs with a learning rate $3\times10^{-5}$ and a batch size 26.
In Figs.~\ref{fig4}g, h of the main text, only the first 15 epochs are displayed due to the loss output retaining the first three decimal places.
In the $16^{\text{th}}$ epoch, some models exhibit training set loss $<$ 0.001.

For the final MgBERT model, we continued training with the highest classification accuracy model weight of Method 3 at a learning rate of $2\times10^{-5}$ and compared it to the case with $1\times10^{-5}$.
The highest accuracy of 88.5\% was achieved when the learning rate was $2\times10^{-5}$, and the epoch was 4.
We utilized the model weights trained under these parameters as the final MgBERT model.

\textit{\textbf{Model evaluation:}}
During the model evaluation stage, we compared three ML models -- their concepts are outlined in the ``Glossary''.
The model implementation is based on the ``scikit-learn'' software and employs its default hyperparameters for training.
The random seed is fixed at 42 for reproducibility.
The training and test set rely on stratified sampling, as described in the ``Methods'' section of the main text.

\begin{algorithm}[H]
  \caption{MgBERT Position Embedding}
  \label{mgembed}
  \begin{algorithmic}[1]
    \State \textbf{Input:} New maximum position embedding $n$, Current position embeddings $E$
    \State \textbf{Output:} Resized position embeddings $E_{new}$
    \State Set $E_{new} = \text{Embedding}(n, E.size(1))$
    \State Set $l = E.shape[0]$
    \State Set $L = \lceil n / l \rceil$
    \If{$l > n$}
      \State Set $E_{new}.weight = E[:n].detach()$
    \Else
      \For{$i \in [0, L-1]$}
        \If{$i == 0$}
          \State Set $E_{new}.weight[:l] = E.detach()$
        \ElsIf{$i > 0 \text{ and } i < L - 1$}
          \State Set $E_{new}.weight[i*l:(i+1)*l] = E.detach()$
        \Else
          \State Set $E_{new}.weight[i*l:] = E.detach()[:(n - i*l)]$
        \EndIf
      \EndFor
    \EndIf \\
    \Return $E_{new}$
  \end{algorithmic}
\end{algorithm}

\subsubsection{Model interpretability}
We used \emph{LIME}~\cite{ribeiro2016should} for model interpretation.
\emph{LIME} utilizes a trained local surrogate model to explain individual predictions.
For a black-box model of interest, \emph{LIME} samples around the instance of interest and collects perturbed instances and their black-box model predictions.
A local interpretable model (e.g., linear regression, decision tree) is then trained on this new dataset to approximate the behavior of the black-box model locally.
Fig.~\ref{figS15} from~\cite{ribeiro2016should} shows its working principle, wherein the dashed line is a trained linear model used to explain the black-box model with complex boundaries in the red and blue part of the graph.
We set  hyperparameters to sample 1,000 cases and filter the top 50 contributing features.
Considering the maximum token length does not exceed 900, this choice is reasonable under the limitation of computational rate, as the training time increases exponentially with  increasing sampling times.

Additionally, we visualized the attention score of the [CLS] (classify) token in the last layer.
Different attention scores represent the degree of importance a specific token places on other tokens, with larger attention scores indicating higher importance.
The classifier processes the value of the last layer [CLS] token, compressing this 768-dimensional vector into three dimensions; each dimension represents the probability of different categories.
Therefore, calculating the focus on the previous layer assists in interpreting the model.

\clearpage
\section{Supplementary Figures}

\begin{figure*}[ht]
\centering
\includegraphics[width=0.6\textwidth]{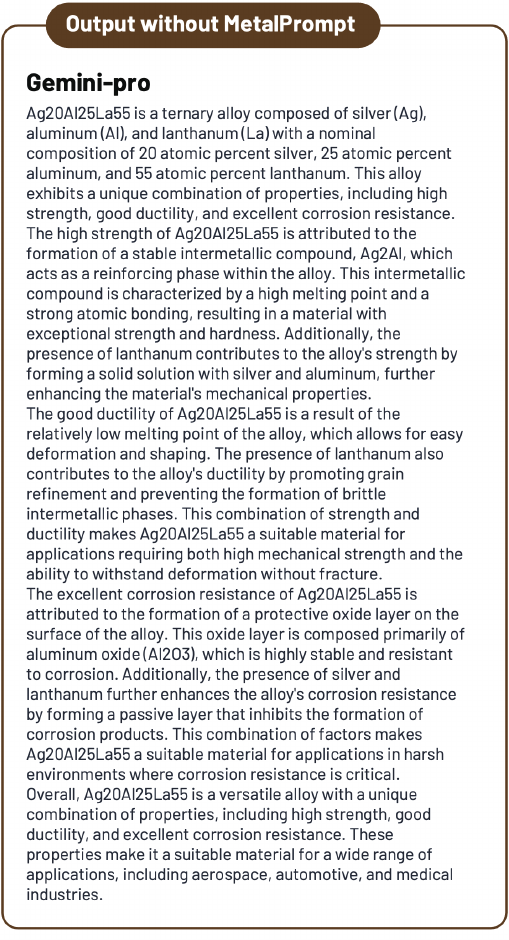}
\caption{\emph{Gemini-pro} outputs without using ``MetalPrompt''.}\label{figS1}
\end{figure*}

\newpage
\begin{figure*}[ht]
\centering
\includegraphics[width=0.6\textwidth]{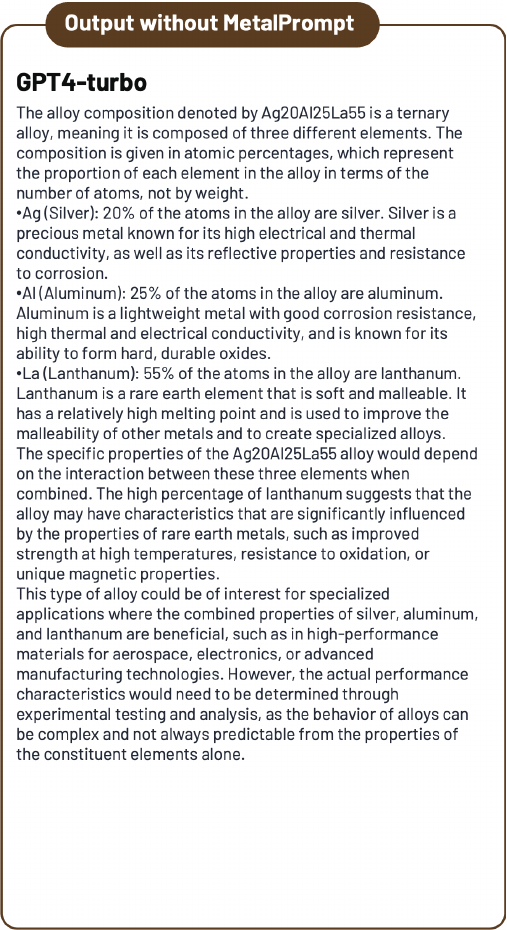}
\caption{\emph{GPT4-turbo} outputs without using ``MetalPrompt''.}\label{figS2}
\end{figure*}

\newpage
\begin{figure*}[ht]
\centering
\includegraphics[width=0.6\textwidth]{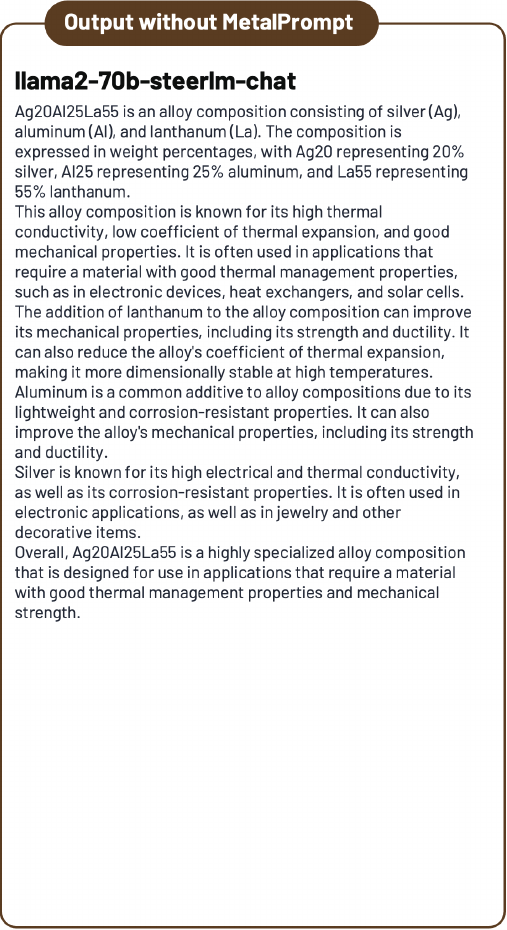}
\caption{\emph{llama2} outputs without using ``MetalPrompt''.}\label{figS3}
\end{figure*}

\newpage
\begin{figure}[ht]
\centering
\includegraphics[width=0.7\textwidth]{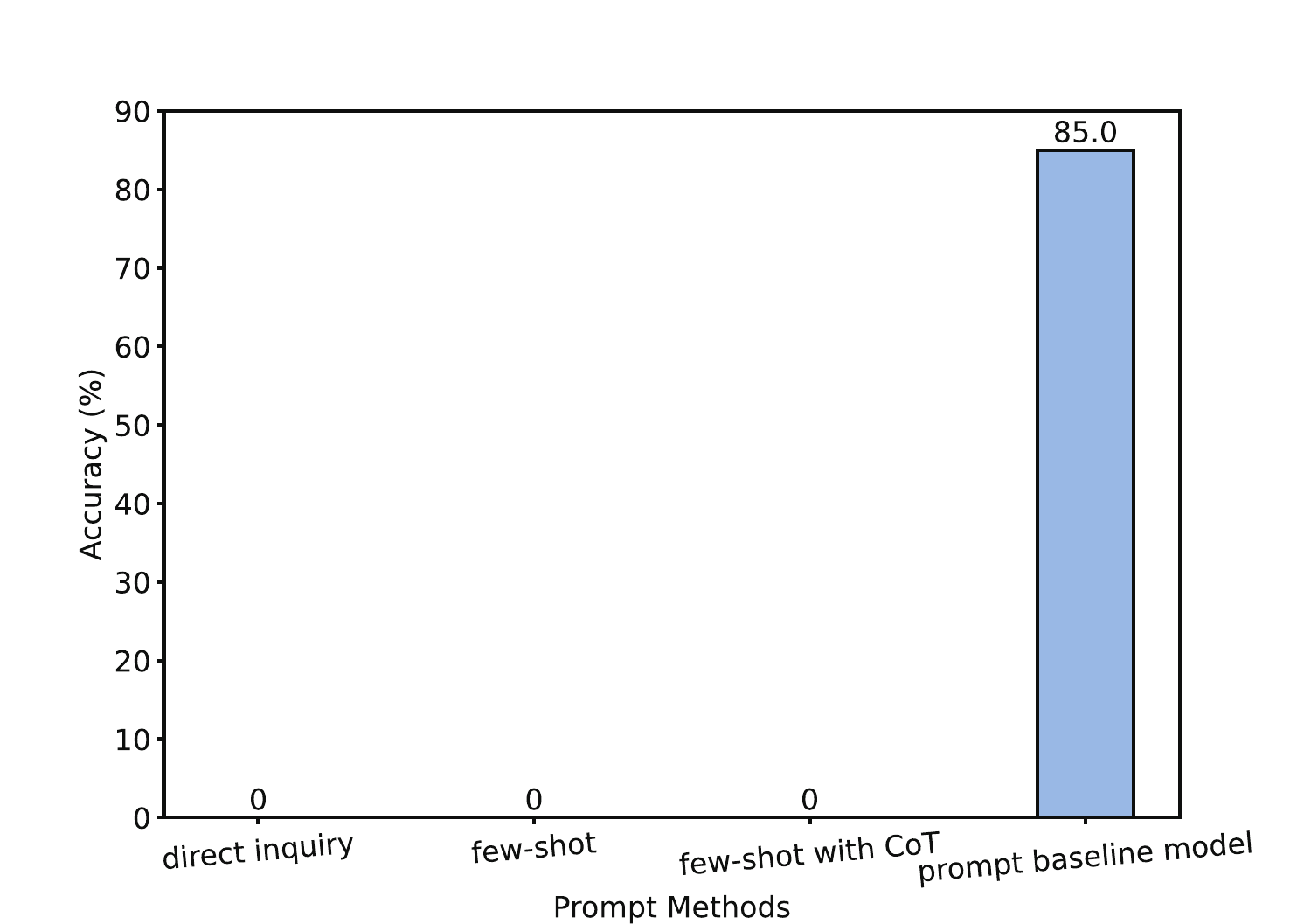}
\caption{Comparison of accuracy of different prompt methods. ``direct inquiry'' refers to directly requesting LLM metallic glass classification results without using a prompt. ``few-shot'' involves using alloy composition and its classification results as input samples in a prompt for an LLM and inquiring about the classification results for other specific alloy compositions. ``few-shot with CoT'' not only adds examples of alloy composition and categories, but also integrates some thought processes into the prompt.}\label{figS4}
\end{figure}

\begin{figure}[ht]
\centering
\includegraphics[width=0.7\textwidth]{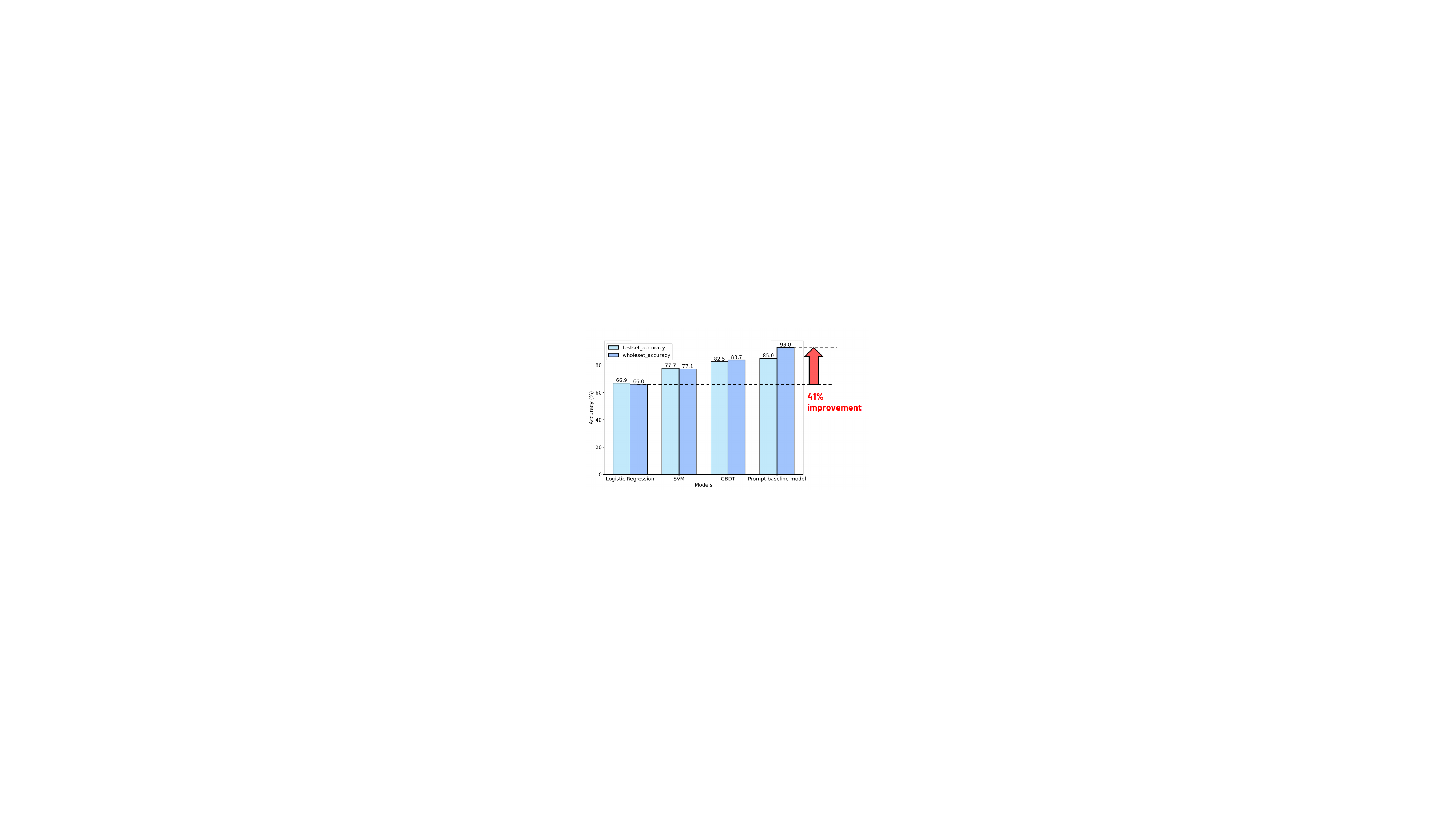}
\caption{Accuracy comparison between prompt baseline model and machine learning models. The logistic regression serves as our baseline model, SVM denotes the support vector machine model, and GBDT refers to the gradient boosting decision tree model.}\label{figS5}
\end{figure}

\newpage
\begin{figure}[ht]
    \centering
    \includegraphics[width=1.0\textwidth]{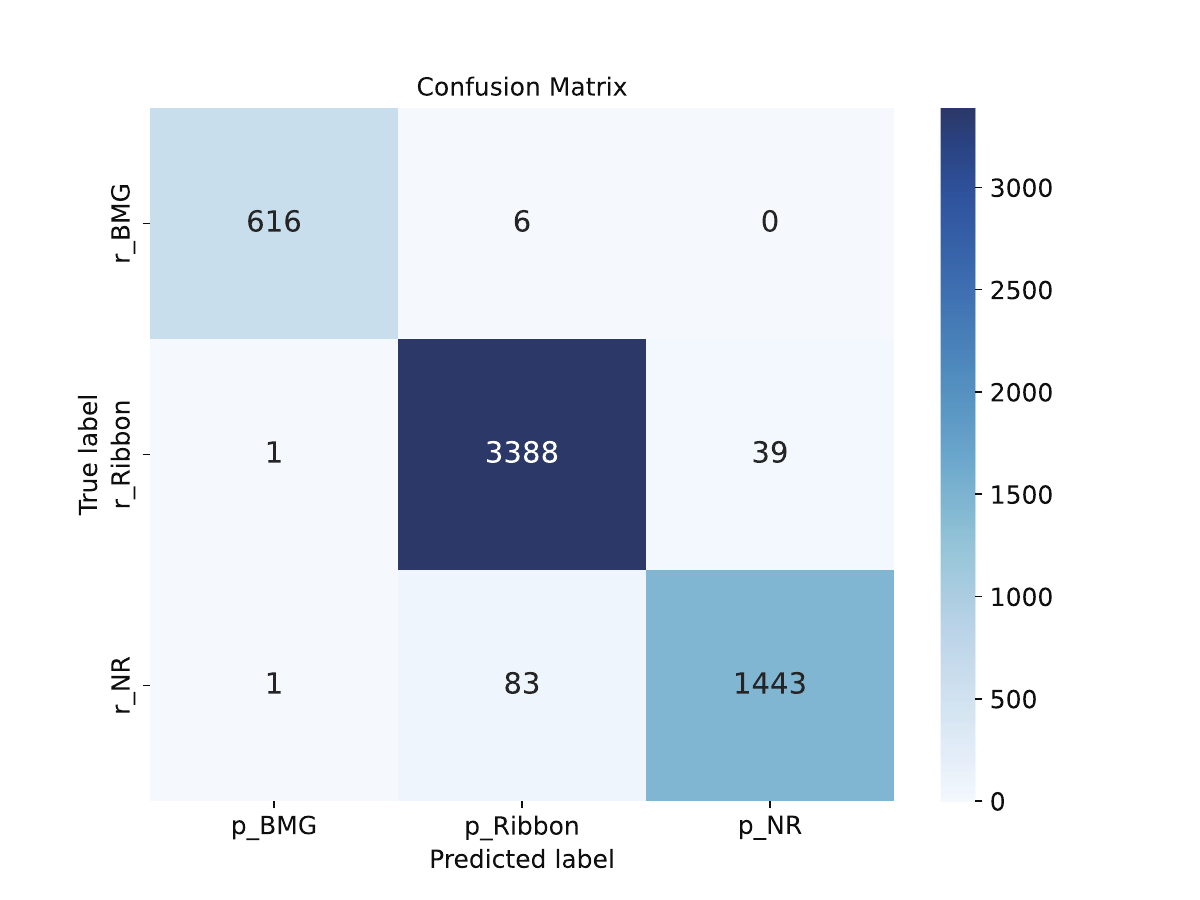}
    \caption{Confusion matrix for MgBERT classification results. The $x$-axis refers to the model prediction of the alloy composition belonging to a particular metallic glass category, while the $y$-axis indicates the actual category of the alloy composition. For example, the `616' in the upper left corner signifies that are 616 components classified as metallic glasses, which MgBERT also predicts to be metallic glasses.}\label{figS6}
    \end{figure}

\newpage
\begin{figure*}[ht]
    \centering
    \includegraphics[width=1.0\textwidth]{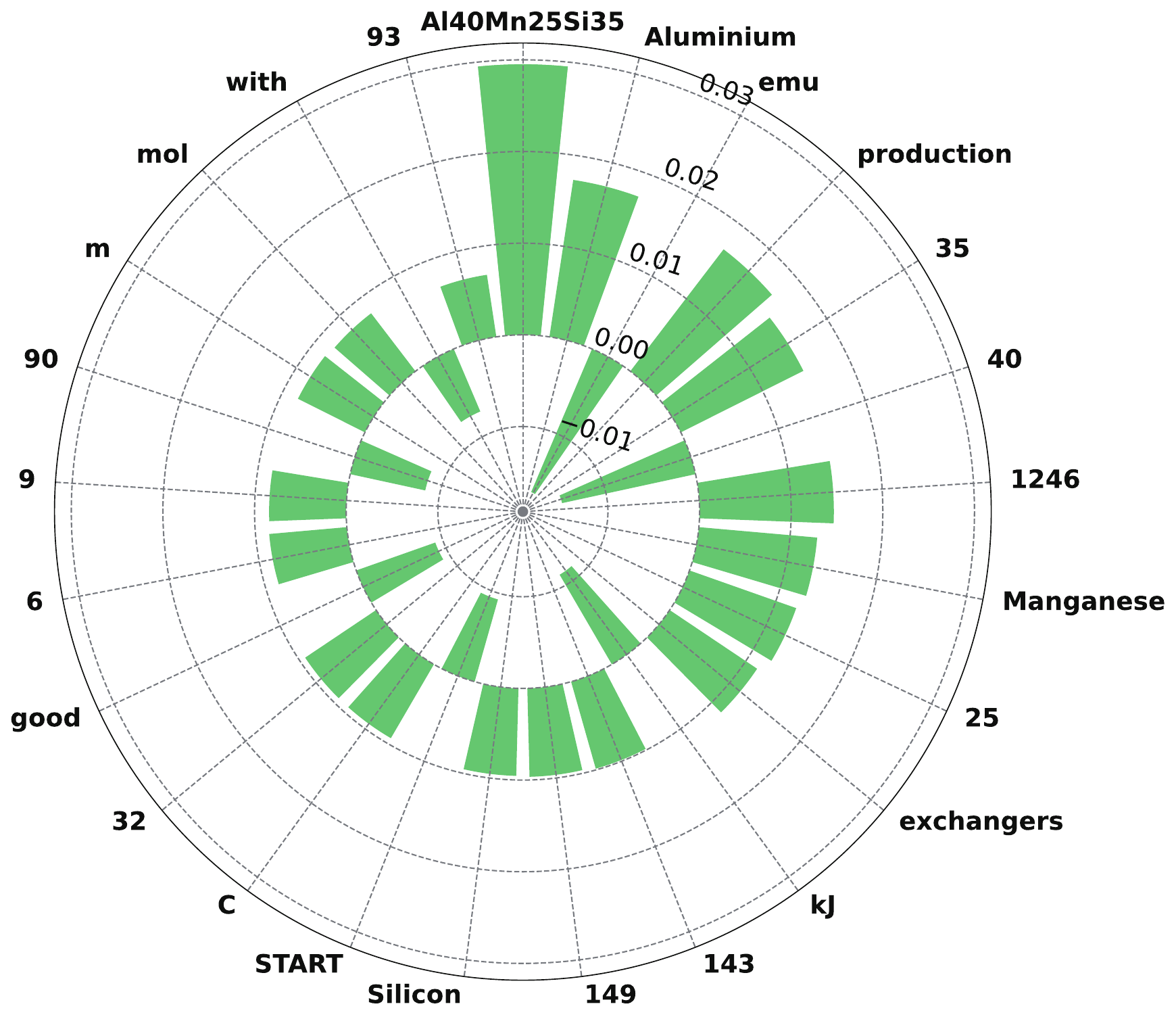}
    \caption{Contribution of input text for non-ribbon type alloy Al$_{40}$Mn$_{25}$Si$_{35}$ to output classification results. A positive value means that the word has a positive contribution to the model's prediction that the component is non-ribbon, and vice versa.}\label{figS7}
    \end{figure*}

\newpage
\begin{figure*}[ht]
    \centering
    \includegraphics[width=0.7\textwidth]{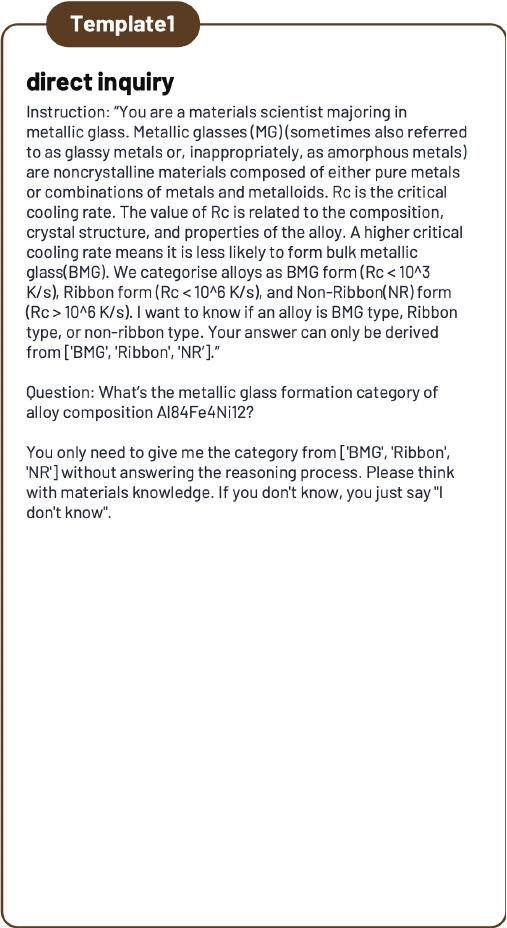}
    \caption{Prompt template for direct inquiry.}\label{figS8}
    \end{figure*}

\newpage
\begin{figure*}[ht]
    \centering
    \includegraphics[width=0.7\textwidth]{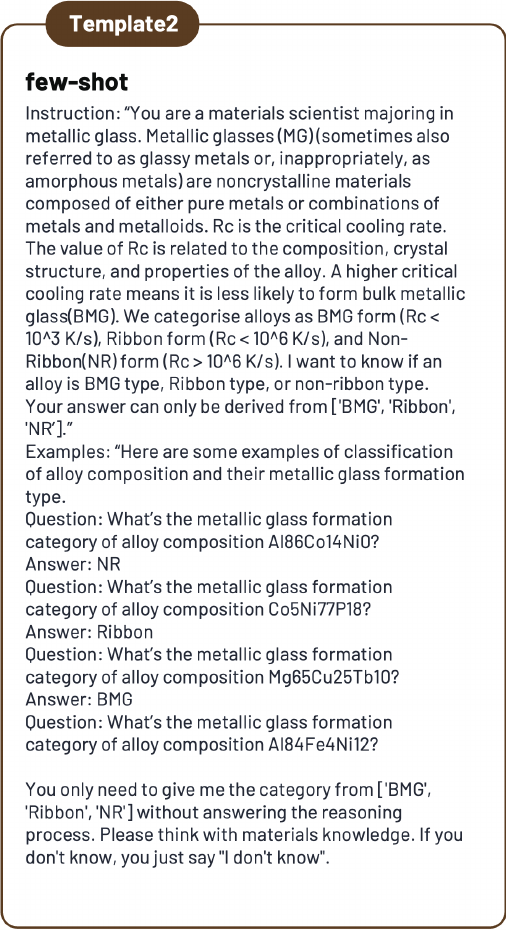}
    \caption{Prompt template for few-shot method.}\label{figS9}
    \end{figure*}

\newpage
\begin{figure*}[ht]
    \centering
    \includegraphics[width=0.8\textwidth]{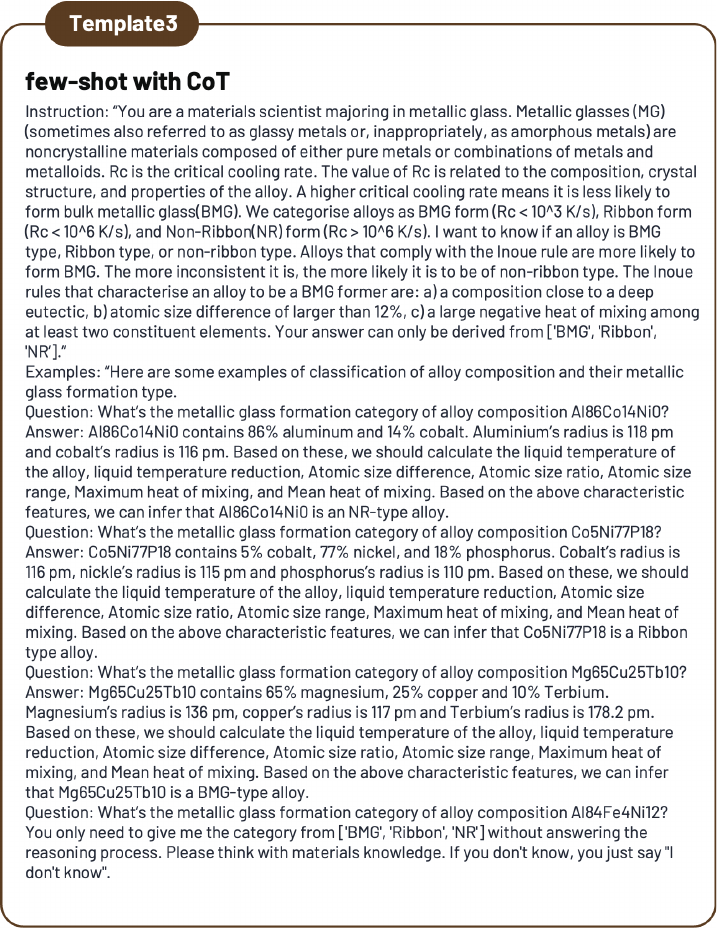}
    \caption{Prompt template for few-shot with CoT method.}\label{figS10}
  \end{figure*}

\newpage
  \begin{figure*}[htbp]
    \centering
    \includegraphics[width=1.0\textwidth]{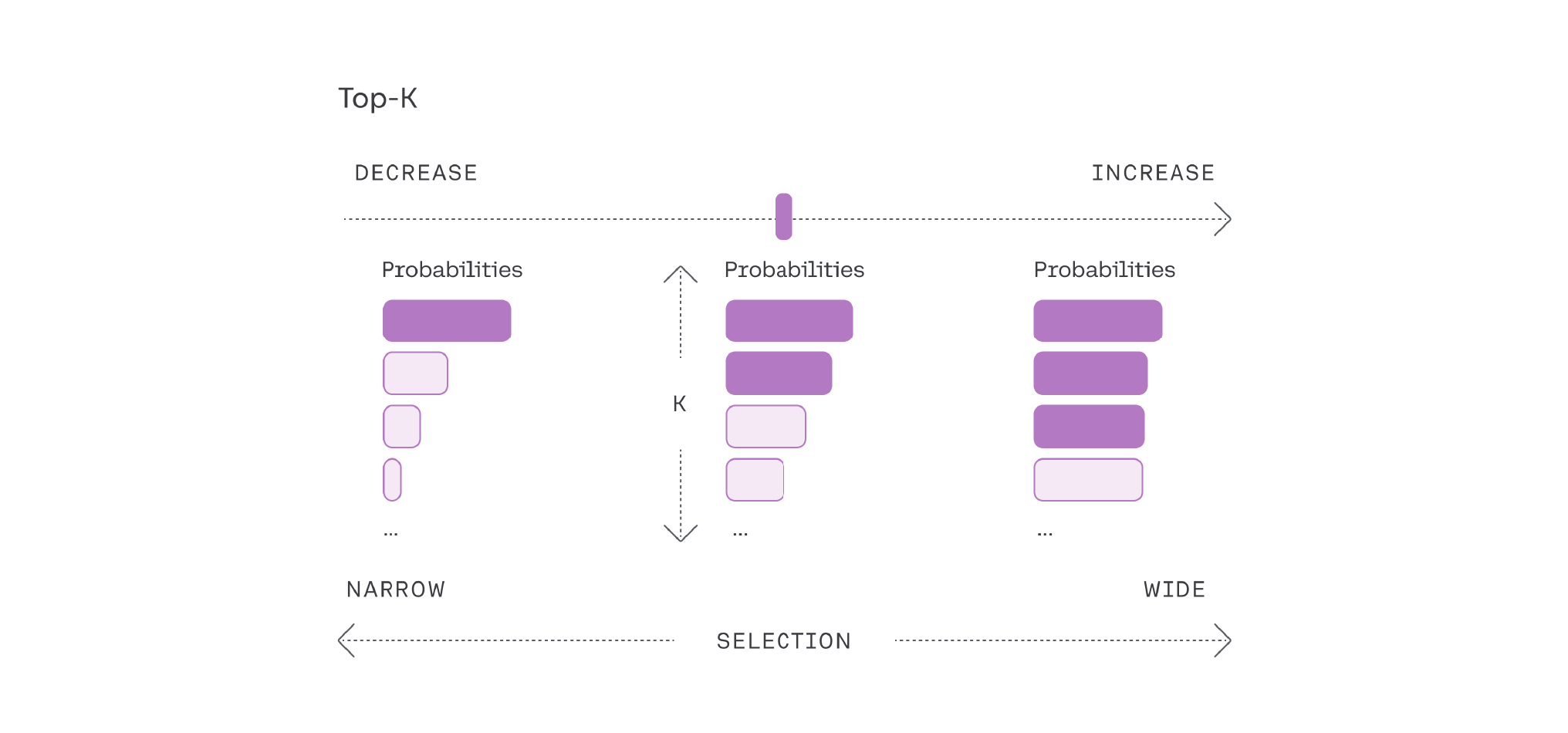}
    \caption{Tuning the Top-K setting~\cite{topk}. The value of K represents the model's selection of the first k words with the highest likelihood of being the output. Decreasing the hyperparameter ``Top-K'' results in a narrower model selection, focusing on candidate words with higher probabilities. This figure is from~\cite{topk}.}\label{figS11}
  \end{figure*}

  \begin{figure*}[htbp]
    \centering
    \includegraphics[width=1.0\textwidth]{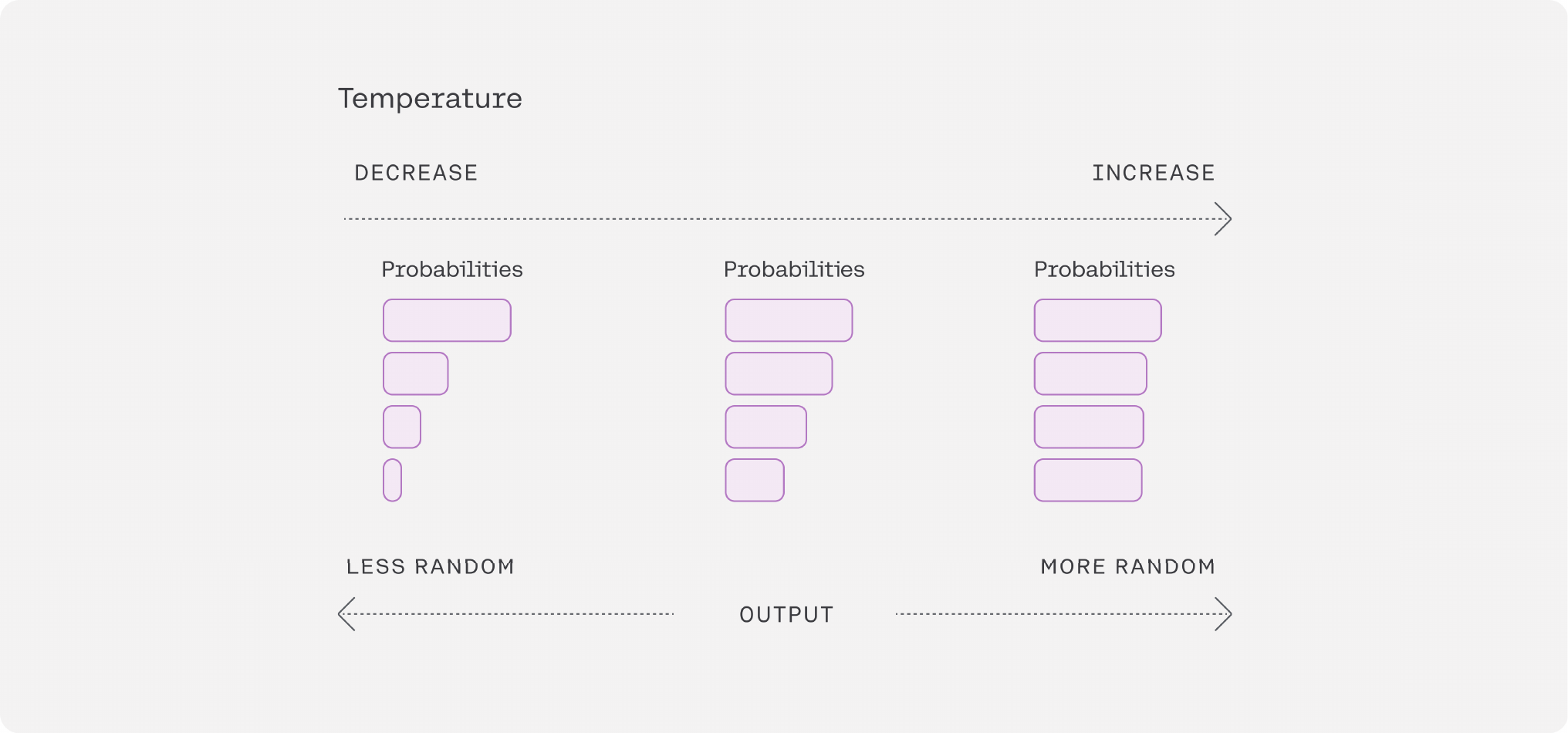}
    \caption{Tuning the temperature setting~\cite{temp}. Similar to the ``Top-K'', a lower value of the hyperparameter temperature implies that the model will favor words with higher probabilities as output. This figure is from~\cite{temp}.}\label{figS12}
  \end{figure*}

\newpage
  \begin{figure*}[htbp]
    \centering
    \includegraphics[width=1.0\textwidth]{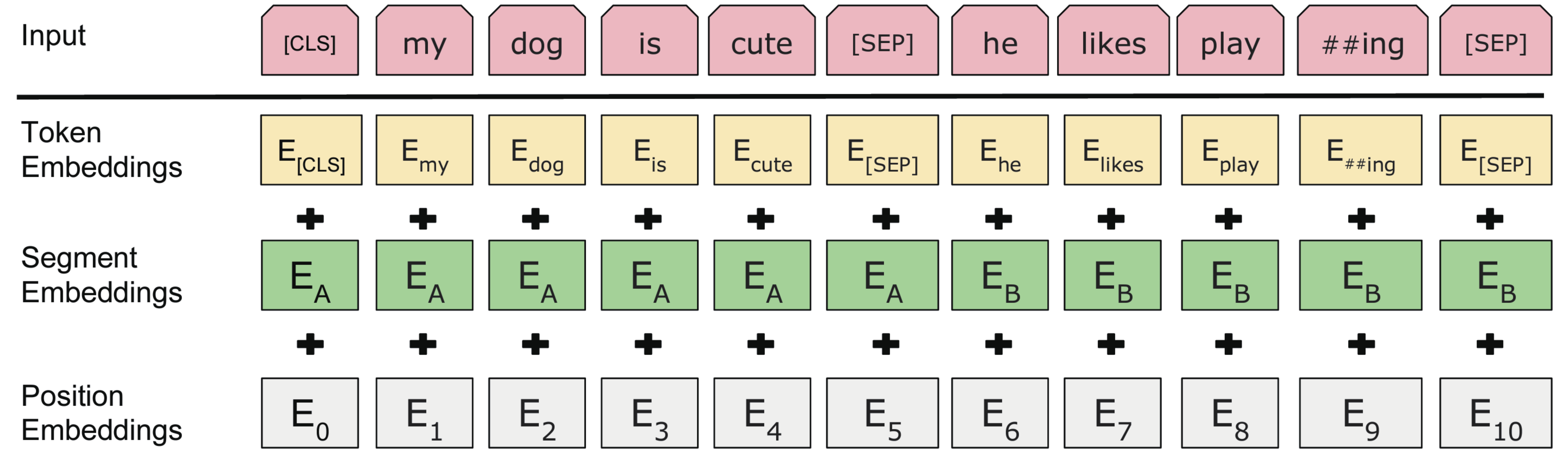}
    \caption{BERT input representation. The input embeddings are the sum of the token, segment, and position embeddings~\cite{devlin2018bert}. This figure is from~\cite{devlin2018bert}.}\label{figS13}
  \end{figure*}

  \begin{figure*}[htbp]
    \centering
    \includegraphics[width=0.8\textwidth]{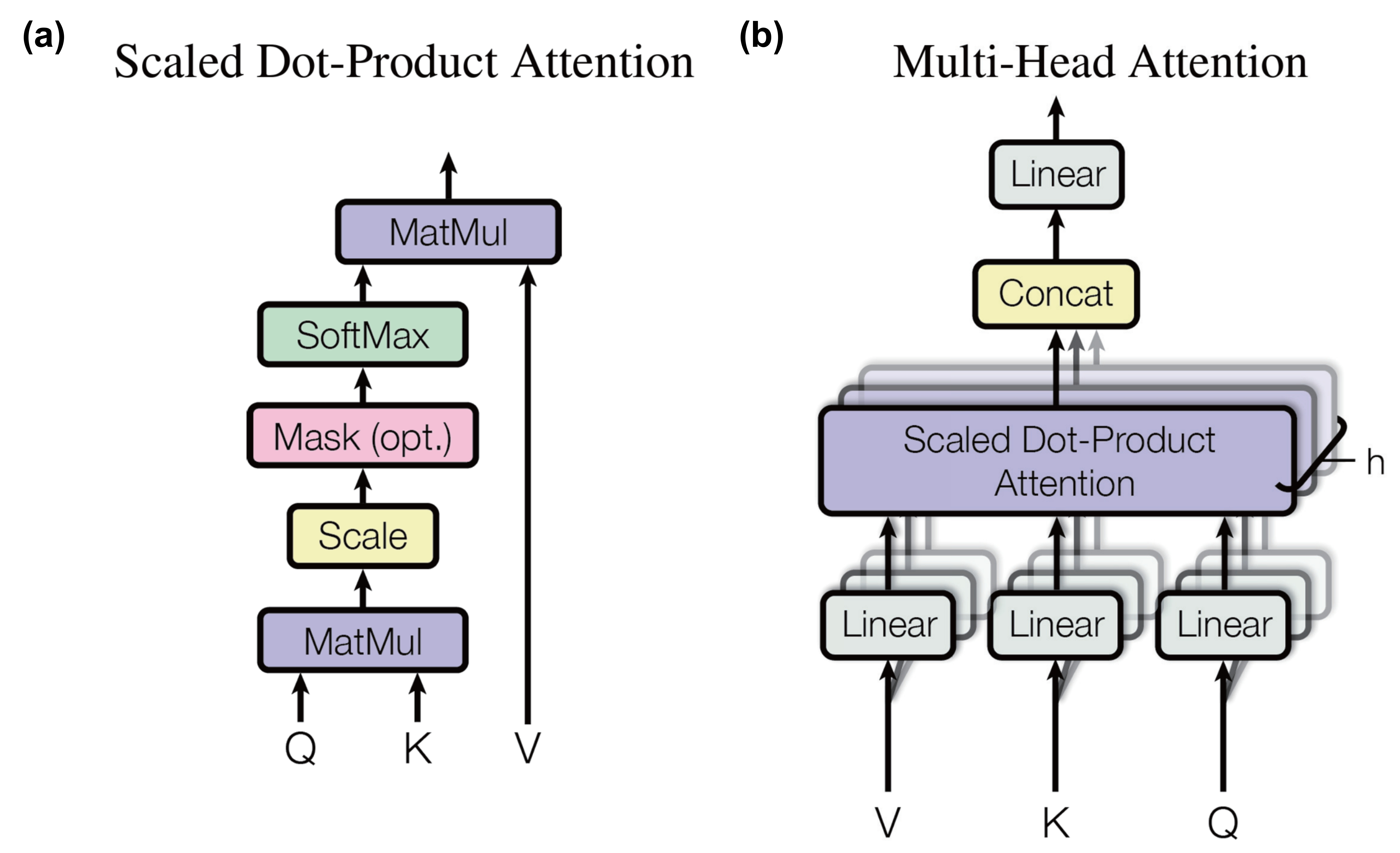}
    \caption{Schematic diagram of attention mechanism. (a), Scaled dot-product attention. (b), Multi-head attention consists of multiple attention layers running in parallel~\cite{vaswani2023attention}. This figure is from~\cite{vaswani2023attention}.}\label{figS14}
  \end{figure*}

  \clearpage

  \begin{figure*}[h!]
    \centering
    \includegraphics[width=0.8\textwidth]{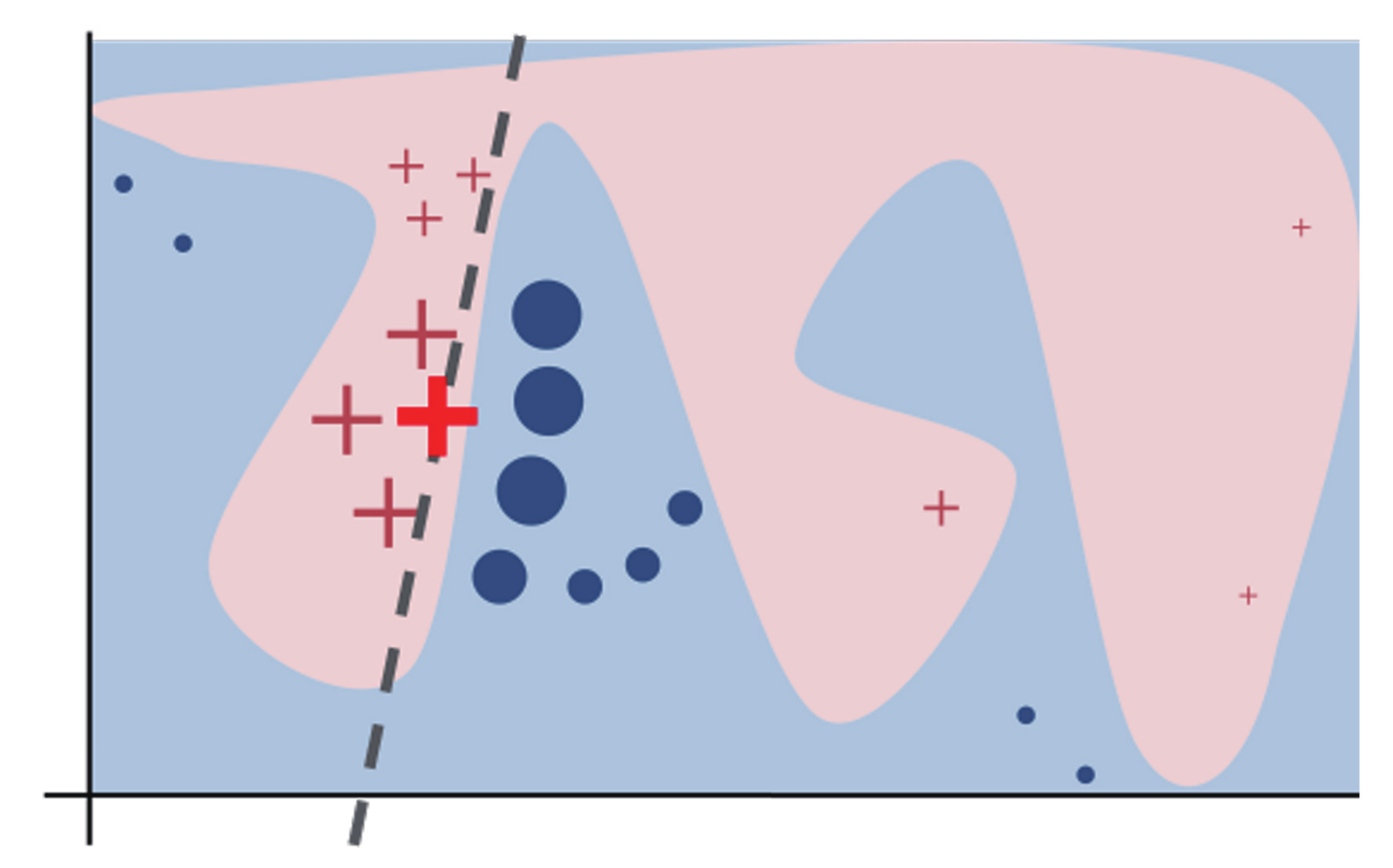}
    \caption{A simple example of \emph{LIME}~\cite{ribeiro2016should}. The gray dashed line is a linear interpreter identified by \emph{LIME}, used to distinguish the positive and negative contributions of sample features to the output of the black box model (i.e., our model). The symbols in the figure correspond to data points generated by \emph{LIME}'s perturbation sampling, while the red and blue areas denote the complex decision areas of the black box model. This figure is from~\cite{ribeiro2016should}.}\label{figS15}
  \end{figure*}

\FloatBarrier

\bigskip

\section{Supplementary Tables}


\begin{table}[h!]
  \caption{Format of raw data.}\label{tableS1}
  \centering
   \begin{tabular}{m{4.0cm} m{6.0cm}}
    \toprule
    glass forming category & composition \\
    \midrule
    Ribbon & Ag20Al25La55 \\
    BMG & Cu55Zr42.5Ga2.5 \\
    NR & Ag6Ce8Cu86 \\
    ... & ... \\
    \bottomrule
  \end{tabular}
\end{table}

\begin{table}[h!]
  \caption{Hyperparameters of model training.}\label{tableS2}
  \centering
  \begin{tabular}{m{4.0cm} m{6.0cm}}
    \toprule
    model name & hyperparameters \\
    \midrule
    BERT & Epoch: 8; Learning rate: 3e-5; Batch size: 26 \\
    Longformer & Epoch: 8; Learning rate: 3e-5; Batch size: 12 \\
    MatSciBERT & Epoch: 8; Learning rate: 3e-5; Batch size: 26 \\
    \bottomrule
  \end{tabular}
\end{table}

\end{document}